\documentclass{ws-ijmpa}
\usepackage{url}
\usepackage[super,compress]{cite}
\usepackage{slashed} 
\usepackage{nccmath} 
\usepackage{times}
\usepackage{amsmath} 
\usepackage{amssymb} 
\usepackage{verbatim} 
\usepackage{graphicx} 
\usepackage{color} 
\usepackage{booktabs} 
\usepackage{empheq}
\usepackage{relsize}
\usepackage{fancyref}
\usepackage{xcolor,colortbl}
\definecolor{olive}{rgb}{0.3, 0.4, .1}
\definecolor{fore}{RGB}{249,242,215}
\definecolor{back}{RGB}{51,51,51}
\definecolor{title}{RGB}{255,0,90}
\definecolor{dgreen}{rgb}{0.,0.6,0.}
\definecolor{gold}{rgb}{1.,0.84,0.}
\definecolor{JungleGreen}{cmyk}{0.99,0,0.52,0}
\definecolor{BlueGreen}{cmyk}{0.85,0,0.33,0}
\definecolor{RawSienna}{cmyk}{0,0.72,1,0.45}
\definecolor{Magenta}{cmyk}{0,1,0,0}
\definecolor{lcyan}{rgb}{0.6,1,1}
\usepackage[usenames,dvipsnames,pdf]{pstricks}
\usepackage{epsfig}
\usepackage{pst-grad} % For gradients
\usepackage{pst-plot} % For axes
\usepackage{pstricks-add}
\usepackage[off]{auto-pst-pdf} % for revtex, compile by hand and switch this line off
\newcommand{\nths}{\negthickspace\negthickspace} 
\newcommand{\nthn}{\negthinspace\negthinspace}
\newcommand{\lp}{\left(} \newcommand{\rp}{\right)} 
  
\newcommand{\ls}{\left[}  \newcommand{\rs}{\right]}
\newcommand{\lv}{\left|}  \newcommand{\rv}{\right|}

\newcommand{\cd}{\!\cdot\!}
\newcommand{\st}[1]{\slashed{#1}}
\mathchardef\mhy="2D   % define short minus sign
\DeclareMathOperator{\Ai}{Ai}
\DeclareMathOperator{\J}{J}

\renewcommand{\labelitemi}{$\bullet$}

\begin{document}
\markboth{Anthony Hartin}{Strong field QED in lepton colliders and electron/laser interactions.}

\catchline{}{}{}{}{}
%
%%%%%%%%%%%%%%%%%%%%%%%%%%%%%%%%%%%%%%%%%%%%%%%%%%%%%%%%%%%%%%%%%%%%

\title{Strong field QED in lepton colliders and electron/laser interactions.}

\author{Anthony Hartin}

\address{University College London, Gower Street, \\London WC1E 6BT, U.K.}

\maketitle

\begin{history}
\received{5 April 2018}
\revised{16 April 2018}
\end{history}

\begin{abstract}
Studies of strong field particle physics processes in electron/laser interactions and lepton collider interaction points are reviewed. These processes are defined by the high intensity of the electromagnetic fields involved and the need to take them into account as fully as possible. Thus, the main theoretical framework considered is the Furry interaction picture within intense field quantum field theory. In this framework, the influence of a background electromagnetic field in the Lagrangian is calculated non perturbatively, involving exact solutions for quantised charged particles in the background field. These "dressed" particles go on to interact perturbatively with other particles, enabling the background field to play both a macroscopic and microscopic role. Macroscopically, the background field starts to polarise the vacuum, in effect rendering it a dispersive medium. Particles encountering this dispersive vacuum obtain a lifetime, either radiating or decaying into pair particles at a rate dependent on the intensity of the background field. In fact, the intensity of the background field enters into the coupling constant of the strong field quantum electrodynamic Lagrangian, influencing all particle processes. A number of new phenomena occur. Particles gain an intensity dependent rest mass shift that accounts for their presence in the dispersive vacuum. Multi photon events involving more than one external field photon occur at each vertex. Higher order processes which exchange a virtual strong field particle, resonate via the lifetimes of the unstable strong field states. Two main arenas of strong field physics are reviewed; those occurring in relativistic electron interactions with intense laser beams, and those occurring in the beam beam physics at the interaction point of colliders. This review outlines the theory, describes its significant novel phenomenology and details the experimental schema required to detect strong field effects and the simulation programs required to model them. 

\end{abstract}

\ccode{PACS numbers: 11.10.St, 12.20.Ds, 12.20.Fv, 12.38.Lg}

\newpage\tableofcontents\newpage

\section{Introduction}	

\begin{figure}
\centering
\includegraphics[width=0.38\textwidth]{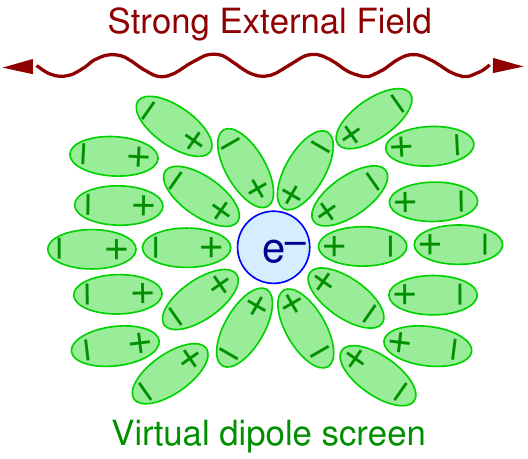}
\caption{\bf\label{fig:frog2}Vacuum dipole screening and re-alignment in a strong external field.}
\end{figure}

We know that the vacuum plays a key role in our physical theories, nevertheless our understanding of the physical vacuum is still relatively limited. In classical physics, the vacuum is merely the unchanging background in which physical phenomena occur. However the quantum vacuum is understood to consist of virtual particles, many of which are charged. An external electromagnetic field can couple to these virtual charges, which in turn affects the behaviour of real particle processes (figure \ref{fig:frog2}). \\

When a strong electromagnetic field is present, the virtual charges, which form virtual dipoles, start to separate under the influence of the field \cite{GreRei02}. In the Schwinger limit, where an electric field ($1.3 \!\times\! 10^{18}$ V/m) does the work equivalent to separating two rest masses over a Compton wavelength, the vacuum state becomes unstable and the field is predicted to induce vacuum pair production \cite{Dunne09}. \\ 

%\begin{wrapfigure}{L}{0.4\textwidth}

Strong fields that reach the Schwinger limit are present in ultra relativistic heavy ion collisions, where the Coulomb field is altered by 0.1\% on a scale of the Compton wavelength \cite{Rafel98}. Strong background fields are also present in an astrophysical setting near the surface of a magnetar \cite{DauHar83}. Strong gravitational fields sufficient to produce pairs are present near a black hole \cite{Hossen13} or during the period of cosmological inflation \cite{Martin08}. \\ 

High power laser facilities \cite{ELI16,Vulcan06} are also a key strong field arena, in which ultra high field intensities are provided by laser pulses. These facilities have renewed modern interest in strong field phenomena \cite{Panek03a,HarHeiIld09,BocFlo09,MacDiP11,SeiKam11,Hartin11a,Hartin15,DiPiaz12,Heinz12}.\\

Strong field physics theory also predicts observable, novel resonant phenomena at field strengths well below the Schwinger limit \cite{Hartin17a}. When the strong field is provided by a laser field, the key parameter is a ratio of the energy provided by the strong field to the relativistic mass of the electron, denoted by $\xi$. The intensity parameter $\xi$ reaches unity at an intensity of $I\approx 10^{19}$ W cm$^{-2}$ for a 1 $\mu$m wavelength laser. These intensities are already produced routinely in the laboratory, with several facilities planned which will well exceed $\xi=1$. \\

Another domain in which strong background fields are evident, is the interaction points (IP) of particle colliders. The quest for new particle physics discoveries in current and future experiments requires very high luminosities which lead to demanding beam conditions. These beam conditions lead to large beam beam effects in which the intense field of the squeezed charge bunches approach the Schwinger critical field in the rest frame of incoming particles. These strong fields are unavoidable at the IP, and all physics processes at the IP occur within them. The precision requirements for future linear colliders demand an equally precise understanding of these strong field effects (section \ref{sect:colliders}). \\

The theoretical framework often used to study strong field physics is intense field quantum field theory (IFQFT). In this framework, the background field is treated non perturbatively, appearing in the parts of the Lagrangian describing charged particles, rather than the interaction term. IFQFT changes our physical picture, making the background field plus vacuum a dispersive medium in which particles that couple to it, decay. This new vacuum enables resonant transitions for propagating particles parameters that correspond to transitions between quasi-energy levels (section \ref{sect:theory}). \\ 

Preliminary studies show that these resonant transitions are likely to be detected in electron/laser interactions with currently available technology. Theoretical calculations that describe these resonant transitions have made much progress, but there is still work to be done. The IFQFT renormalisation procedure is yet to be fully established. This requires a non standard treatment, since the electron spin couples both to the self energy and the background field. Also, the possibility of an unstable vacuum at high enough field intensities, requires special treatment\cite{FraGitShv91,GavGit18}. \\

Phenomenologically, the location and widths of resonant, differential transition probabilities can be calculated (section \ref{sect:res}) and included in a full simulation taking into account experimental realities such as charged bunch dynamics and laser pulse shape (section \ref{sect:sim}). Experimentally, schemas can be prepared to discover these predicted resonances, as well as other novel strong field effects, leading to sensitive tests of IFQFT (section \ref{sect:exp}). A successful outcome will shed light on our understanding of the vacuum, its dispersive nature in the presence of fields, and predict phenomenology for quantum field theories in strong background potentials in general. \\

IFQFT physics processes can be sorted into first order and higher order processes with respect to their Furry picture Feynman diagrams. High intensity Compton scattering (HICS) and one photon pair production (OPPP) are the first order processes that have attracted much study (section \ref{sect:elelas}). Just as interesting are the higher order processes with a strong field propagator such as stimulated Compton scattering (SCS) and stimulated two photon pair production (STPPP). These higher order processes exhibit resonance behaviour and promise to be a potentially important experimental tool in uncovering new physics (section \ref{sect:res}). \\

There already exist several review works on aspects of strong field physics. References [\refcite{Ritus79,Nikishov79}] provide a thorough review of extensive theoretical work done since the 1960s up to 1980. [\refcite{Hartin06}] has an introduction on strong field QED processes concentrating equally on first and higher order processes, while [\refcite{DiPiaz12,EhlKam09}] give a general review of strong field laser/electron studies, concentrating on recent work. [\refcite{Ugger05}] reviews strong field physics in crystal channels in which the strong fields are of a constant, crossed form. Simulation of strong field processes in real interactions require an approach based on particle in cell (PIC) methods which are reviewed in [\refcite{Gonos15}]. \\

As good as review papers, are several PhD theses dedicated to strong field processes. An examination of the background theory and first order processes in laser pulses was performed by [\refcite{Mackenroth14}]. The influence and phenomenology of strong field processes in a laser pulse was extended to the case of two photon Compton scattering \cite{Seipt12}. A thorough investigation of the higher order resonances in two vertex Compton scattering and pair production was carried out by [\refcite{Hartin06}]. Two works on the strong field beam beam processes at a linear collider interaction point proved invaluable for collider studies \cite{Schulte96,Schroeder90}. \\

This review paper will give a summary of the IFQFT theory, some of the exact solutions that exist for that theory, and will touch upon the issues involved in more comprehensive approaches. The first order processes will be reviewed, but some current topics of interest, such as radiation reaction, cascades and laser field breakdown will not be covered. Instead, the review concentrates on specific higher order processes produced in relativistic electron/intense laser interactions, including experimental signatures. The equivalent processes occurring in the beam beam interaction at the interaction point of a future linear collider will also be surveyed. A comprehensive simulation program which can cope with the range of phenomena presented will also be described. \\

To the extent that analytic work is presented, a metric with a (1,-1,-1,-1) signature will be used. Natural units will be utilized and the Lorenz gauge will be chosen.
%The Lorenz gauge requires scalar products of external field 4-momentum and 4-potential to vanish, $k.A^e=0$ which prove useful in obtaining alternative forms for the Volkov solutions and strong field fermion propagator, 

\begin{align}\label{eq:notn}
\text{metric: }&\quad g_{\mu\nu}=(1,-1,-1,-1) \notag\\
\text{units: }&\quad c=\hbar=4\pi\epsilon_0=1, \quad e=\sqrt{\alpha} \\
\text{gauge: }&\quad \partial^\mu A^\text{e}_\mu(k\cd x)=0 \implies k\cd A^\text{e}=0 \notag
\end{align}

%Light cone coordinates which rotate the hyperplane containing the time coordinate (t) and the propagation direction of the external field (z) will be denoted by $\pm,\perp$ subscripts. So light-cone scalar products can be written 

%\begin{gather}
%p\cd x=\vec{p}_\text{-}\cd\vec{x}_\text{+}+p_\text{+}\,x_\text{-},\,\, \vec{p}_\text{-}\!\equiv\!\{p_\text{-},p_\perp\},\,\, \vec{x}_{+}\!\equiv\!\{x_\text{+},x_\perp\} \\
%p_\text{-}\equiv \epsilon_p-p_\text{z},\quad p_\text{+}=\mfrac{p_\perp^2+m^2}{2p_\text{-}} \notag
%\end{gather}

\section{Historical development}\label{sect:history}

The historical evolution of strong field physics has occurred over a long period originating near the beginning of quantum physics itself. A review article by [\refcite{Dunne12}] covers the Euler-Heisenberg effective action and its effect on quantum field theory. The Klein paradox and subsequent work is described by [\refcite{NarFed15}]. [\refcite{Schweber62}] has a substantial section on the Furry picture (FP) and its origin at the birth of QED. \\
 
The foundation of strong field quantum field theory lies in it's understanding of the quantum vacuum. The modern outlook began from the time of the Dirac equation and the conception of a Dirac sea of virtual particles and anti-particles. These virtual charges were understood to respond to an applied electromagnetic field, and consequently the Dirac equation, minimally coupled to an electromagnetic field, was solved soon after \cite{Volkov35}. \\

The strange nature of the quantum vacuum was apparent in the Klein paradox when calculations showed that the electron wave would penetrate a large potential barrier and carry negative energy \cite{Klein29,Sauter31,Hund27,Hund41}. An important consequence of Klein's paradox was the creation of electron positron pairs from the vacuum by extended, constant electrostatic fields\cite{HeiEul36}. One direction this work lead to was strong field effects in heavy ion collisions\cite{Greiner85}. \\

The Furry picture arose from attempts to calculate the Lamb shift at the end of the 1940s. The first relativistic treatments were non covariant and the treatment of divergences was ambiguous \cite{Schweber62}. Moreover, within perturbation theory the effective coupling constant for experiments with high Z atoms ($Z\alpha$) lead to non-convergence of the perturbation series. \\

[\refcite{Furry51}] solved both of these problems by including the strong field non perturbatively within a Lagrangian treatment of quantum electrodynamics\cite{Feynman48a,Feynman48b,Tomonaga46,Tomonaga47,Tomonaga48,Schwinger48a,Schwinger48b}. Since a canonical transformation linked field operators in the Heisenberg and Schr\"odinger pictures to the new formulation, Furry's treatment was termed the Furry picture. The historical terminology of the bound Dirac equation and the bound interaction picture also had it's origin in the need to calculate the Lamb shift at that time\cite{LamRut47}. \\

In the early 1960s, the theoretical treatment of processes within polarised, oscillatory fields \cite{Reiss62,BroKib64,NikRit64a,NikRit64b,ReiEbe66} was spurred on by the experimental development of the LASER \cite{Einstein17,SchTow58,Maiman60}. An alternative, relativistic quantum mechanics approach to strong field physics proved suitable for relativistic particles \cite{Baier68,Baier69,Baier72,Baier75,Baier09}. \\

Theoretical work continued from the 1960s in a number of directions. In one development of much interest in this review, it was realised that strong field propagators could reach the mass shell and lead to resonant cross sections for a series of kinematic conditions \cite{Oleinik67,Oleinik68,Oleinik72,Bos79a,Bos79b,Hartin06,Roshchup96}. Analytic calculations of strong field processes was aided with the formulation of IFQFT on the light cone \cite{NevRoh71,NevRoh71b,KogSop70}. The formulation of IFQFT for field strengths exceeding the Schwinger limit in which the vacuum itself became unstable, was also an important development \cite{Fradkin81,Fradkin88,FraGitShv91}. \\

In recent times, a landmark experiment at SLAC which produced pairs from the vacuum with a laser assisted trident process was planned and carried out \cite{McDonald86,McDonald89,McDonald91,Bamber99}. This experiment prompted a flurry of modern work on the strong field QED processes that will be described in the ongoing sections of this review. First, a brief overview of the theoretical framework of IFQFT will be given. 

\section{Theoretical strong field physics}\label{sect:theory}

Strong field physics will be considered here as a non perturbative quantum field theory. Though classical treatments of strong field phenomena exist, the QFT treatment is more general, converging with the classical theory in appropriate limits. Indeed, there are several phenomena which only emerge with a full quantum treatment. The reason for this lies in the nature of the physical vacuum within which strong field physics occurs.\\

We begin with an exposition of the Furry interaction picture (section \ref{sect:furry}). The solutions of the Dirac equation in a plane wave electromagnetic field are given, as are the Green's function solutions for strong field propagators (section \ref{sect:Volkov}). The form of electromagnetic fields expected in laser and collider interactions is defined in section \ref{sect:fields}. Later, the virtual sector of the theory is examined by considering the Furry picture self energies and the regularisation of strong field propagators (section \ref{sect:selfene}).

\subsection{The Furry picture and the Strong field QED Lagrangian}\label{sect:furry}

\begin{comment}
\begin{figure}[b]
%\centering\begin{subfigure}[t]{0.5\textwidth}
\centerline{\includegraphics[width=0.6\textwidth]{/home/hartin/Physics_Research/myfigs/1storder_vertex.pdf}}
\caption{\bf One vertex HICS process.}\label{fig:hics}\vspace{0.1cm}
%\end{subfigure}\begin{subfigure}[t]{.5\textwidth}
\end{figure}\begin{figure}
\centerline{\includegraphics[width=0.5\textwidth]{/home/hartin/Physics_Research/myfigs/scs_directchannel_xy.pdf}}
\caption{\bf Two vertex SCS process.}\label{fig:scs}
%\end{subfigure}\caption{\bf Feynman diagrams for 1st and 2nd order IFQFT processes}
\end{figure}
\end{comment}

The Furry picture is a type of interaction picture used to deal with interactions between fermions and bosons in the presence of an intense electromagnetic field, such as that provided by a focussed laser or the charge bunch fields at the interaction point of a collider \cite{Furry51,Schweber62,JauRoh76,Hartin11a}. This framework treats the intense field as a background, calculates the exact wave function for the electron embedded in that background \cite{Volkov35,BagGit90}, quantises those solutions and then interacts them with other particles. \\

If the external field is sufficiently strong, such that quantum interactions with it leave it essentially unchanged, the external field can be considered to be classical. The fermion interaction with the external potential can be calculated exactly by including it in the Dirac part of the Lagrangian density, meaning that the Dirac field operators $\psi^\text{FP}$ will differ from their usual free field selves. The QED Lagrangian density in the Furry picture, containing the quantized fermion $\psi^\text{FP}$ and photon $A$ fields as well as the external classical field $A^e$ is,

\begin{align}\label{eq:furpic}
\mathcal{L_{\text{QED}}^{\text{Furry}}}&=\bar\psi^\text{FP}(i\slashed{\partial}\!-\!e\slashed{A}^e\!-\!m)\psi^\text{FP}-\textstyle{\frac{1}{4}}(F_{\mu\nu})^2-e\bar\psi^\text{FP}\!\slashed{A}\,\psi^\text{FP} 
\end{align}

Treating the final term in the Lagrangian density of equation \eqref{eq:furpic} as the interaction term in the usual perturbation theory, the Euler-Lagrange equations for the $A$ and $\psi^\text{FP}$ fields in the remainder of the Lagrangian density, lead to the usual equation of motion for the gauge boson field, and a bound Dirac equation, minimally coupled to the background field $A^\text{e}$,

\begin{gather}
(i\slashed{\partial}\!-\!e\slashed{A}^e\!-\!m)\psi^\text{FP}=0, \quad\partial^2 A=0
\end{gather}

The state of the vacuum in the Furry picture $|\Omega^e\rangle$ is an issue in at least two respects. If the external field is strong enough, an electron can be lifted from the Dirac sea across the energy gap between negative and positive energy solutions to leave the vacuum charged. This means that the vacuum expectation value of the electromagnetic current must no longer vanish and tadpole terms in the S-matrix expansion contribute \cite{Schweber62}

\begin{gather}
\lv\langle \Omega^e|\bar\psi^V\gamma_\mu\psi^V|\Omega^e\rangle\rv^2 \neq 0
\end{gather}

Secondly, the second quantization of the Dirac field relies on there being a gap between negative and positive energy solutions. Since this gap closes by the action of the external field, the separation into creation and destruction operators is problematic. This point, where the gap between negative and positive energy solutions vanishes, is considered to be the limit of the validity of the Furry picture. Beyond it, one must employ a procedure that deals with an unstable vacuum \cite{FraGitShv91}. \\

Generally, and in almost all the work presented in the rest of this review, it is assumed that the the strength of the background field is less than the Schwinger critical field\cite{Schwinger54}, so that the question of an unstable vacuum doesn't arise. Nevertheless, the unstable vacuum bears closer examination and remains a fascinating subject in its own right.

\subsection{Dirac equation solutions in a plane wave, background electromagnetic field}\label{sect:Volkov}

Initially, the minimally coupled Dirac equation was solved exactly (the Volkov solution), when the external field consists of plane waves and a single propagation direction \cite{Volkov35}. Other authors extended this initial work, with [\refcite{Sengupta67}] solving the Dirac equation for an electron in an external field consisting of two polarised plane electromagnetic waves. [\refcite{Bagrov74,Bagrov75}] discussed the exact solution of a relativistic electron interacting with a quantised and a classical plane wave travelling in the same direction, and [\refcite{Fedorov75}] proposed a method of constructing a complete orthonormal system for the electron wave function for an electron embedded in a quantised monochromatic electromagnetic wave. \\

Exact solutions exist also for Coulomb fields, longitudinal fields, pulsed fields, and fields of colliding charge bunches \cite{BagGit90,BocFlo09,HeiSeiKam10,SeiKam11,Hartin15}. [\refcite{Varro14}] has extended the class of solutions for charged particles travelling in a medium, while exact solutions in pair creating electric fields is reviewed in \cite{GavGit17}. \\

For an electron of momentum $p_\mu\!=\!(\epsilon,\vec{p})$, mass m and spin r embedded in a plane wave electromagnetic field of potential $A^e_{\text{x}\mu}$ and momentum $k_\mu=(\omega,\vec{k})$, and with normalisation $n_\text{p}$ and Dirac spinor $u_\text{rp}$, the Volkov solution is,

\begin{gather}\label{eq:Volkov}
 \Psi^\text{FP}_\text{prx}= n_\text{p}\,E_\text{px}\; u_{\text{pr}}\;e^{- i p\cdot x },\quad n_\text{p}=\sqrt{\mfrac{m}{2\epsilon(2\pi)^3}}\, \\
E_\text{px}\equiv\ls 1 - \mfrac{\slashed{A}^e_\text{x}\st{k}}{2(k\cd p)}\rs e^{-i\mathlarger{\int}^{k\cdot x}\;\mathlarger{\frac{2eA^{e}_\xi\cdot p - e^2A^{e\,2}_\xi}{2k\cdot p}}d\xi},\quad A^\text{e}_\text{x}\equiv eA^\text{e}(k\cd x), \quad A^\text{e}_\xi\equiv eA^\text{e}(\xi) \notag
\end{gather} 

When the external field is provided by a laser field (with a period of oscillation $2\pi L$), the Volkov solution can be expanded in a Fourier series of modes corresponding to momentum contributions $nk$ \cite{NarNikRit65},

\begin{gather}\label{eq:Volkov}
 \Psi^\text{FP}_\text{prx}= \sum^\infty_{n=-\infty}\int^{\pi L}_{-\pi L} \mfrac{d\phi}{2\pi L} \; n_\text{p}\,E_{\text{p}\phi}\; u_{\text{pr}}\;e^{- i (p+nk)\cdot x }\;
\end{gather} 

The fermion solutions in the external potential are orthogonal and complete \cite{BerVar80a,BocFlo10,Filip85} and can be quantized with wave packets formed \cite{NevRoh71}. The LSZ formalism can be extended for the propagation of in and out Volkov states \cite{IldTor13}. For Furry picture interactions, the two point correlation function for Volkov states is also required. \\

The strong field propagator for the electron embedded in the external field proceeds from the Green's function solution $G^\text{FP}_\text{yx}$ for the Dirac equation with a source. The solution is obtained by noting that the canonical momentum operator $\st{\pi}^\text{e}$, operating on the Volkov $E_\text{p}$ function yields the free electron momentum $\st{p}$

\begin{gather}
\ls\st{\pi}^\text{e}-m\rs G^\text{FP}_\text{yx}=\delta^4(x-y),\quad \st{\pi}^\text{e}\equiv i\st{\partial}_x-e\st{A}^\text{e}_\text{x},\quad  \st{\pi}^\text{e}E_\text{py}=E_\text{py}\st{p} \notag\\
\implies\quad G^\text{FP}_\text{yx}=\int\mfrac{\text{d}p}{(2\pi)^4}E_\text{py}\;\mfrac{\st{p}+m}{p^2-m^2+i\epsilon}\;\bar E_\text{px}
\end{gather}

The strong field photon propagator differs from its usual perturbation theory in the denominator where it couples to the external field through it's self energy (section \ref{sect:selfene}). There are now enough elements in place to use Feynman diagrams with Furry picture fermion states to study particular physics processes in strong background fields. \\

\subsection{Electromagnetic fields in strong field environments}\label{sect:fields}

For the two strong field arenas that this review concentrates on, the external electromagnetic fields have two main forms. The first are the constant crossed fields present in relativistic charged particle bunches in colliders. The second are the oscillatory electromagnetic fields associated with intense lasers. \\

Constant crossed fields are the form of relativistic charge bunches in which the spherically symmetric fields associated with charges at rest are relativistically squeezed in the transverse direction by the motion of ultra high energy accelerated charge bunches. The electric field $\vec{E}$ is transverse from the centre of the bunch to the periphery and the magnetic field $\vec{B}$ is azimuthal, so that at any position within the charge bunch, the fields are orthogonal and the 4-potential has a space-time dependence that is a simple scalar product with its 4-momentum, 

\begin{equation}
\vec{E}\cdot \vec{B}=0, \quad A^\text{e}_\mu=a_\mu \,k\cd x 
\label{constcross}\end{equation}

The electromagnetic fields associated with laser beams have usually been treated as oscillatory, infinite plane waves (IPW). This is reasonable for an interaction with a laser pulse, as long as the interaction occurs close to the axis defined by the pulse propagation direction, and the pulse is sufficiently long. \\

In terms of the polarisation state of an IPW, the most general description is in terms of an amplitude $|\vec{a}|$ a combination of two transverse field directions ($e^\mu_1,\,e^\mu_2$), a polarisation parameter $\rho$ and oscillatory functions with a dependence on the field 4-momentum and space-time \cite{Panek02,SeiKam11}

\begin{equation}
A^\text{e}=|\vec{a}|\lp e^\mu_1\,\cos{\rho}\,\cos{k\cd x}+e^\mu_2\,\sin{\rho}\,\sin{k\cd x}\rp,\quad e_1\cd e_2=0
\end{equation}

The polarisation state is determined by the angle $\rho$, with $\rho=0,\pi/2,\pi,3\pi/2$ defining different states of linear polarisation, $\rho=\pi/4,3\pi/4,\pi,5\pi/4,7\pi/4$ defining circular polarisation, and other values for $\rho$ defining elliptical polarisation. \\

Various attempts have been made to extend or go beyond the Volkov solution. In order to account for depletion of the external field it seems appropriate to use solutions for the electron in a quantised external field. This description can be obtained by using coherent states \cite{Glauber63,Glauber63b}. A real laser pulse has a temporal and transverse gaussian shape and significant effort has been made to take the pulse shape into account \cite{HeiIldMar10,MacDiP11,Mackenroth14,SeiKam11,Seipt12,Seipt17}.
Intense laser pulses are almost always focussed. The wavefronts of the associated electromagnetic field and, strong field calculations have attempted to take them into account, though there is still much to do on this front\cite{Hartin91,Derlet95,Harvey16}. \\

To underline the importance of using a realistic representation for the external field, the locally constant field approximation has been recently under review \cite{DiPiaz17,DinTor18}. This approximation apparently holds at very high electromagnetic intensities where strong field processes are formed in a small fraction of the background field phase. Nevertheless, the approximation doesn't hold for some parameter ranges, even at very high intensity. \\

There is enough basic, strong field theory in place to consider specific processes within the Furry picture. The first order processes are reviewed first.

\section{Strong field, electron/intense laser interactions}\label{sect:elelas}

Though strong fields are present in the laboratory during collider interactions, the investigation of strong field effects there is often limited by the messy environment they find themselves in. That is to say, strong field effects in colliders are not the primary aim of such experiments, even though they are important to take into account fully. \\

Instead, precision strong field studies can be planned in dedicated experiments involving interactions between intense lasers and relativistic electron beams. The fields of intense lasers are not the same as those in collider bunch collisions, nevertheless experience gleaned from intense laser interactions informs studies of strong field effects in colliders as well. \\

\begin{figure}[h!]
%\centering\begin{subfigure}[t]{0.5\textwidth}
\centerline{\includegraphics[width=0.7\textwidth]{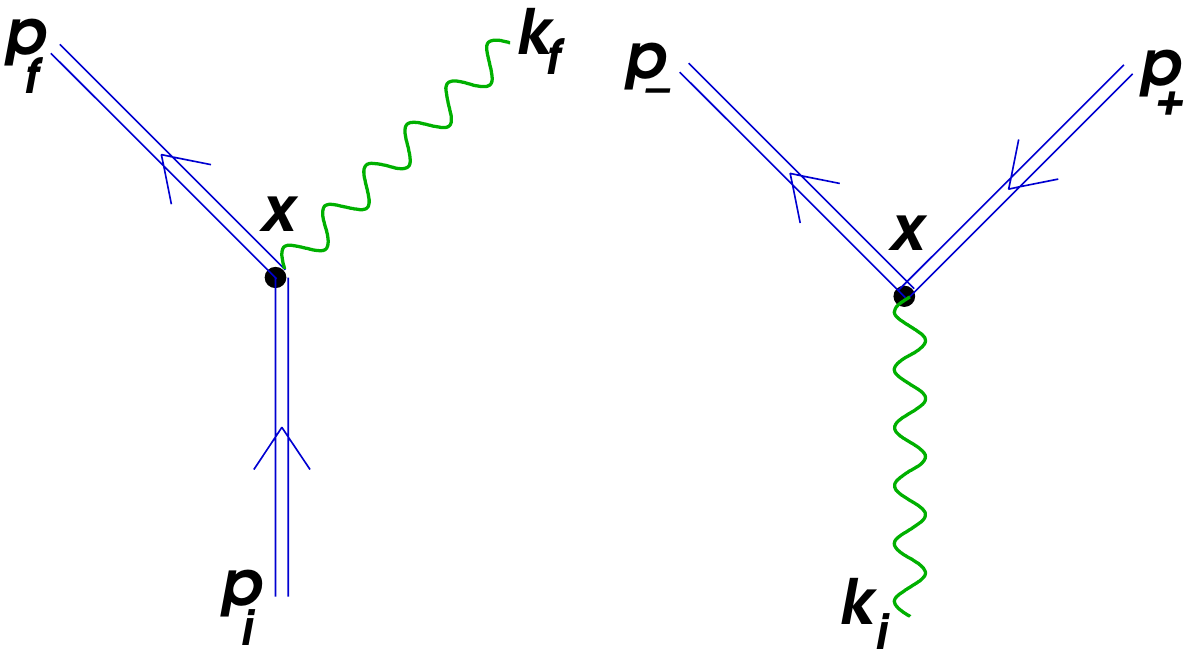}}
\caption{\bf One vertex processes in the Furry picture. High intensity Compton scattering (HICS) on the left, and one photon pair production (OPPP) on the right. Time flows from bottom to top.}\label{fig:hics}\end{figure}

In the normal perturbation expansion of the S-matrix formulation of the QED interaction, the one vertex processes are suppressed. Kinematically, it is usually impossible to produce on-shell particles in one vertex processes. However in the Furry picture, there is extra momentum provided by the external field which makes the one vertex processes possible. Two possible processes - the photon radiation from a bound fermion initial state (HICS process), and the one photon pair production from an initial photon state (OPPP) are studied here (see figure \ref{fig:hics}). \\

%These strong field, one vertex processes, unlike within the normal perturbation series of QED, have a non-zero probability of occurring. This is because the external field provides extra momentum which allows the conservation of 4-momentum for on shell particles in the initial and final states. \\

Formally, these one vertex processes decay due to their coupling to a dispersive vacuum, which was envisaged when the external field was treated non perturbatively in the Lagrangian. Consequently for these processes, decay rates $\Gamma$ are calculated. For this, the standard treatment within say \cite{PesSch95}, section 4.5, which uses relativistic normalisation of the in and out states can be used. For initial particle $(\epsilon_\text{i},\vec{p}_\text{i})$ and a sum over generic final states,

\begin{gather}
\Gamma=\mfrac{1}{2\epsilon_i}\lp\prod_f\mfrac{\text{d}^3\vec{p}_\text{f}}{(2\pi)^3 2\epsilon_f} \rp \lv M_\text{fi}\rv^2\,(2\pi)^4\delta^{(4)}\!\lp p_\text{i}-\sum_f p_\text{f} \rp
\end{gather}

When Furry picture transition probabilities are calculated, for general interactions involving electrons, photons and the external field, three dimensionless parameters characterising strong field interactions appear. These are the intensity parameter $\xi$ related to the field strength of the laser $\vec{E}$, and the electron and photon recoil parameters $\chi,\chi_\gamma$

\begin{gather}
\xi=\mfrac{}{}=\mfrac{e\lv \vec{E} \rv}{\omega m}, \quad \chi=\mfrac{2\xi\,k\cd p}{m^2}, \quad \chi_\gamma=\mfrac{2\xi\,k\cd k'}{m^2}
\label{eq:sqedparams}\end{gather}

Now, specific processes can be studied in detail.

\subsection{High intensity Compton scattering (HICS) in a laser field}\label{sect:hics}

High intensity Compton scattering (HICS) is the primary strong field process in collisions between electron bunches and laser pulses. It has been often studied using classical and semi classical methods. In near head on collisions it is known as inverse Compton scattering and is the main driver of x-ray sources. Here, the process considered will be the one vertex photon radiation in the Furry interaction picture (figure \ref{fig:hics}). \\

Much theoretical work on the HICS process, using the Furry picture and Volkov solutions, was spurred by the construction of the first LASER in 1960 and the possibility of laboratory tests of the theory. The transition amplitude of the HICS process was found to be dependent on the state of polarisation of the external field. For the case of a linearly polarised external electromagnetic field, the HICS transition probability contained an infinite summation of non standard analytic functions which were evaluated in limiting cases only \cite{NikRit64a}. In contrast, a circularly polarised external field results in a transition probability containing Bessel functions, the properties of which are well known \cite{NarNikRit65,BroKib64,Mitter75}. The HICS process for the case of an elliptically polarised external electromagnetic field and for the case of two orthogonal, linearly polarised fields was considered by [\refcite{Lyulka75}].\\

An important effect emerging from the theoretical work on the HICS process was a dependency of the energy of the radiated photon (a frequency shift) on the intensity of the external field \cite{BroKib64,Goldman64}. The existence of this frequency shift allows the HICS process to be used as a generator of high energy photons \cite{Milburn63,Bemporad65,Sandorfi83}. The polarisation properties of high energy photons produced by the HICS process are of fundamental importance in nuclear physics applications, for instance in the study of abnormal parity components in the deuteron wavefunction \cite{GriRek83,RavRam85}. The polarisation state of the emitted HICS photon, as a function of initial polarisation states, for a circularly polarised external field, was studied by [\refcite{GriRek83,Tsai93}].\\

The multi-photon effects in the HICS process were one of the specific, and ultimately successful goals, of the SLAC E144 experiment in the mid 1990s \cite{Bamber99}. The possibility of reproducing and extending this experiment in upcoming electron/laser facilities has prompted several modern studies \cite{SeiKam11,HarHeiIld09,HarHeiIld12,Heinz12}. \\

[\refcite{Hartin11a}] calculated the transition probability for the HICS process for a general external field of any periodicity. Two forms were found, one suitable for an oscillatory field and one for non-oscillatory fields. The HICS calculation itself can be simplified by using an alternative form of the Volkov solution, expressed in the classical action of the electron in the external field $S_\text{px}$, the bispinor $u_\text{rp}$ and a gauge invariant canonical momentum $\Pi_\text{px}$ which combines the free electron momentum $p$ with the external field $A^\text{e}$\cite{Hartin16},

\begin{gather}
\psi_\text{px}=n_\text{p}(\st{\Pi}_\text{px}+m)\mfrac{\st{k}}{2k\cd p}u_\text{rp}\,e^{iS_\text{px}},\quad\Pi_\text{px}=p+k\,\mfrac{2eA^\text{e}_\text{x}\cd p-e^2A^\text{e 2}_\text{x}}{2k\cd p}-eA^\text{e}_\text{x} \\
S_\text{px}=-p\cd x-\medint\int^{k\cdot x}\mfrac{2eA^\text{e}_\phi\cd p-e^2A^\text{e 2}_\phi}{2k\cd p}\,\text{d}\phi\notag
\end{gather}

The HICS transition probability is calculated after simplifying the action. For a circularly polarised infinite plane wave (IPW) laser field of 4-momentum $k$, modes $nk$ are Fourier transformed out, leaving terms containing squares of Bessel functions. For instance,

\begin{gather}
\lv\medint\int\text{d}x\; e^{iS_\text{ix}-iS_\text{fx}-ik_\text{fx}}\rv^2=\sum_n\J^2_n(z_u)\lv\medint\int\text{d}x\;
e^{i\lp p_\text{f}+k_\text{f}-p_\text{i}-nk\rp\cdot x}\rv^2\end{gather}

Then, the integration over final momenta can be performed in the rest frame ($\vec{p}_\text{i}=0$), exploiting the azimuthal symmetry of the circularly polarised external field, leaving one remaining integration over a function of scalar products $u$, 

\begin{gather}
\medint\int\mfrac{\text{d}\vec{k_\text{f}}\;\text{d}\vec{p_\text{f}}}{8\epsilon_\text{i}\epsilon_\text{f}\,\omega_\text{f}}\,\delta^{(4)}\!\lp p_\text{f}+k_\text{f}-p_\text{i}-nk\rp=\mfrac{\pi}{4m}\medint\int \mfrac{\text{d}u}{(1+u)^2},\quad u=\mfrac{k\cd k_\text{f}}{k\cd p_\text{f}}
\end{gather}

Combining the calculation steps and paying attention to the limits of the remaining integration, the transition probability of the HICS process, which appears as a decay rate of the electron in the external field (note that the frame dependence of $\text{d}t$ is balanced by that of $\epsilon_\text{i}$), is

\begin{gather}\label{eq:Wcirc}
\Gamma_{\!\text{HICS}}=-\mfrac{\alpha m^2}{\epsilon_\text{i}}\sum\limits_{n=1}^{\infty}\int_0^{u_n} \nths\mfrac{du}{(1+u)^2} \ls \J_{n}^2-\mfrac{\xi^2}{4}\,\mfrac{1+(1+u)^2}{1+u} \lp\J_{n\text{+1}}^2+\J_{n\text{-1}}^2-2 \J_{n}^2\rp\rs \notag\\
z_\text{u}\equiv \mfrac{m^2\xi\sqrt{1+\xi^2}}{k\cd p_i}\ls u\lp u_n- u\rp\rs^{1/2},\quad u_n\equiv \mfrac{2(k.p_i)\,n}{m^2(1+\xi^2)},\quad \xi\equiv \mfrac{e|A|}{m} 
%W_{\text{HICS}}^{\text{circ}}=-\mfrac{e^2m}{4\pi}\sum\limits_{r=1}^{\infty}\int_0^{u_r} \nths\mfrac{du}{(1+u)^2} \ls 4\J_{r}^2(z)-\xi\,\mfrac{1+(1+u)^2}{1+u} \lp\J_{r+1}^2(z)+\J_{r-1}^2(z)-2 \J_{r}^2(z)\rp\rs \notag\\
%z\equiv \mfrac{m^2}{k\cd p_i}\ls\xi\,u\lp \mfrac{2rk\cd p_\text{i}}{m^2}-(1+\xi)u\rp\rs^{1/2},u_r\equiv \mfrac{2(k.p_i)\,r}{m^2(1+\xi)},\quad \xi\equiv \mfrac{e^2|A|^2}{m^2}
\end{gather}

\begin{figure}[b]
\centerline{\includegraphics[width=0.7\textwidth]{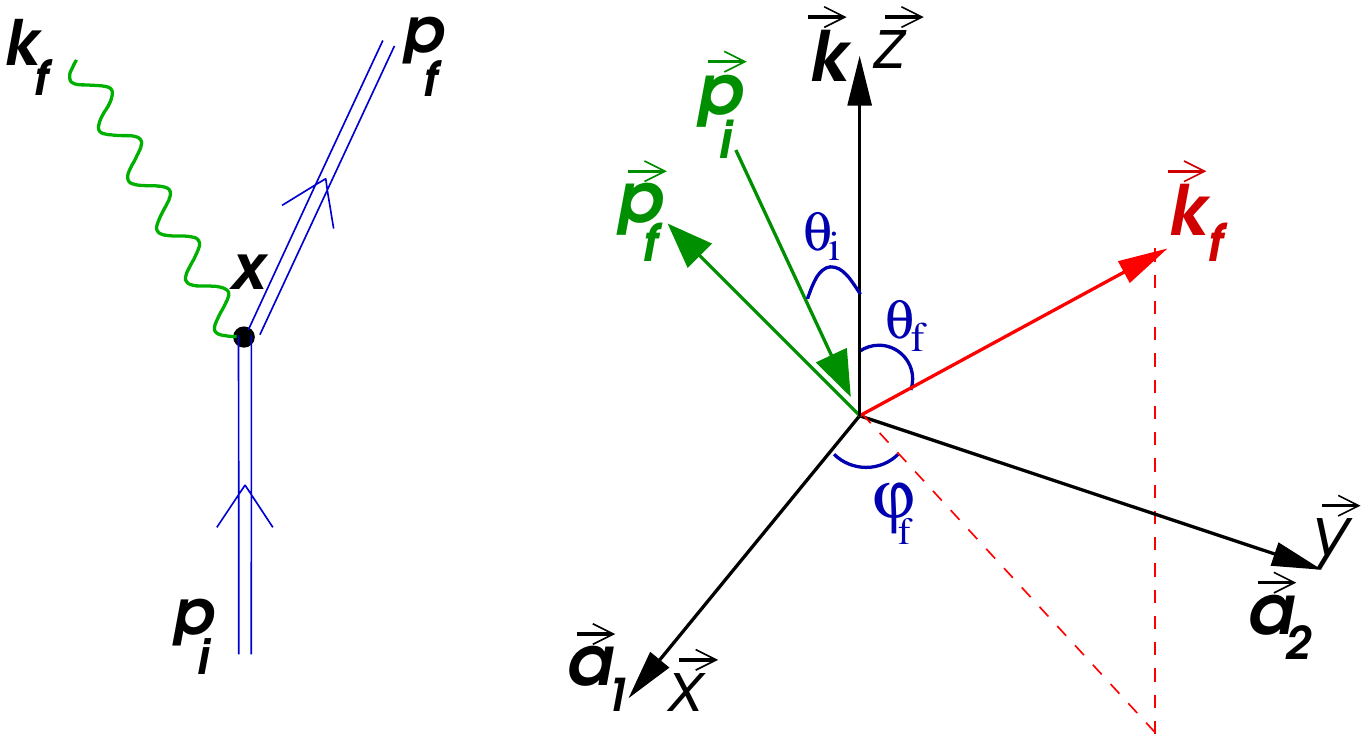}}
\caption{\bf Feynman diagram for high intensity Compton scattering with Volkov electrons $p_\text{i},\; p_\text{f}$. Scattering angles are defined in terms of the coordinate system defined by the external field momentum $\vec{k}$ and potentials $\vec{a}_1,\, \vec{a}_2$. \label{fig1}}
\end{figure}

The remaining phase space integration $\text{d}u$ can be converted to an integration over final photon energy $\omega_\text{f}$ or the radiation angle $\theta_\text{f}$. The propagation direction of the laser also defines the light cone component of the initial electron, $p_{\text{i}-}$ (see figure \ref{fig1}) \\

\begin{gather}
u=\mfrac{k\cd k_\text{f}}{k\cd p_\text{f}},\quad \medint\int_0^{u_n} 
\mfrac{\text{d}u}{(1+u)^2}=\medint\int_0^{\omega_{\text{f}n}} \mfrac{(1-\cos\theta_f)\,\text{d}\omega_\text{f}}{p_{\text{i}-}},\quad \omega_{\text{f}n}=\mfrac{\varepsilon_\text{i}+n\omega}{m}-\sqrt{1+\xi} \notag\\
\medint\int_0^{u_n} 
\mfrac{\text{d}u}{(1+u)^2}=\medint\int_0^{\theta_{\text{f}n}} \mfrac{\omega_f\,\text{d}(1-\cos\theta_\text{f})}{p_{\text{i}-}},\quad \theta_{\text{f}n}=\mfrac{p_{\text{i}-}}{\omega_\text{f}}\mfrac{u_n}{1+u_n},\quad p_{\text{i}-}=\varepsilon_\text{i}-|\vec{p}_\text{i}|\cos\theta_i
\end{gather}

Arising from the conservation of 4-momentum, there is a dependence of the energy of the radiated photon on the radiation angles,

\begin{gather}
\ls p_i+k\,\mfrac{m^2\xi^2}{2k\cd p}-k_f+nk\rs^2=m_\ast^2 \notag\\[4pt]
\omega_f=\mfrac{n\,\omega \,(\epsilon_\text{i}-|\vec{p}_\text{i}|\cos\theta_\text{i})}{\epsilon_i-|\vec{p}_i|\cos\phi_\text{f}\sin\theta_\text{i}\sin\theta_\text{f}-|\vec{p}_i|\cos\theta_\text{i}\cos\theta_\text{f}+(n\omega+m^2\xi^2/2p_{\text{i}-})(1\!-\!\cos\theta_f)}
\end{gather}

\begin{figure}[h!]
%\centering\begin{subfigure}[t]{0.5\textwidth}
\centerline{\includegraphics[width=0.8\textwidth]{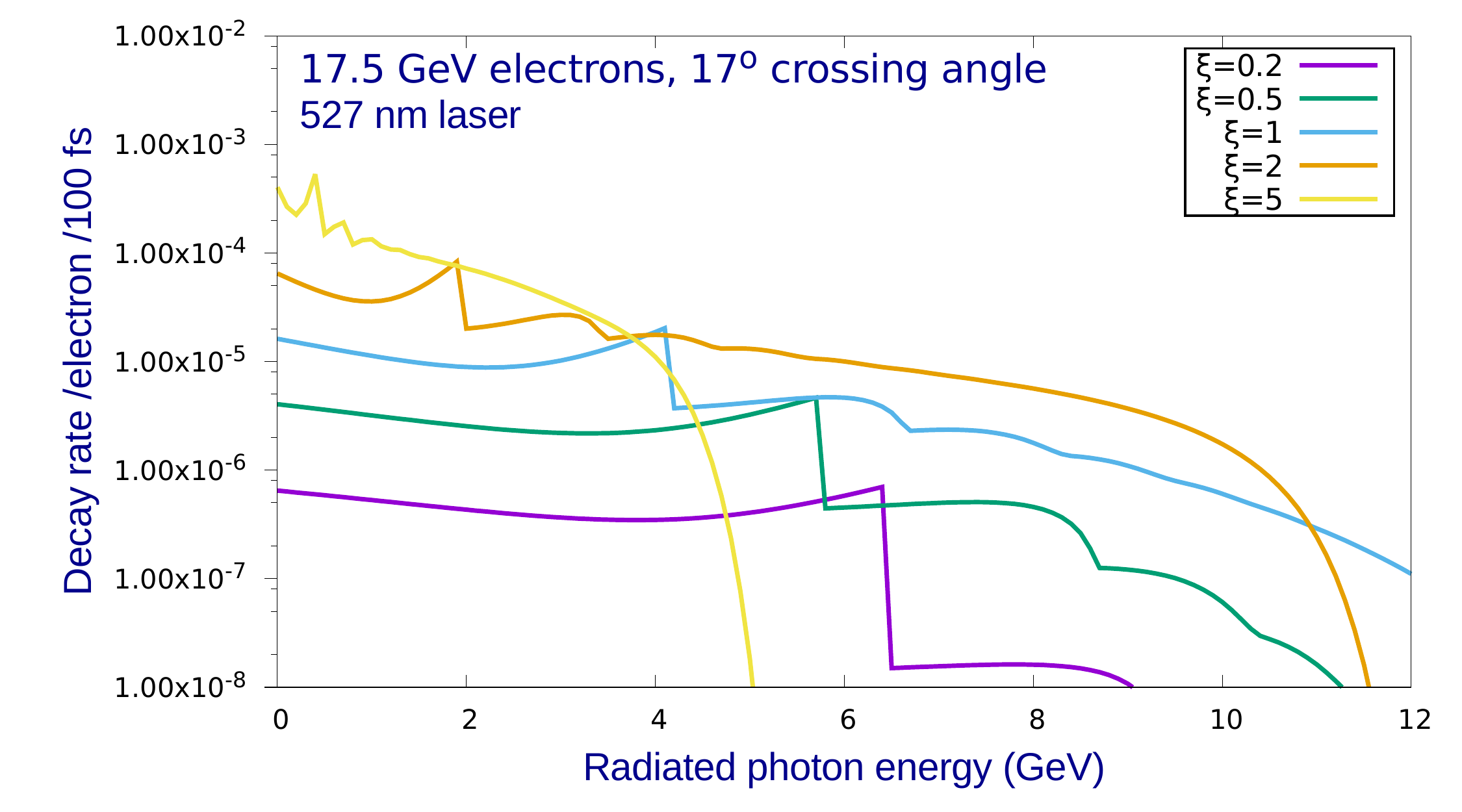}}
\caption{\bf HICS transition rate vs radiated photon energy.}\label{fig:hicscomp}\end{figure}

So, a plot of the analytic transition rate for radiation energy (figure \ref{fig:hicscomp}) can be produced for a set of parameters that can be imagined for a possible future experiment,

\begin{gather}
\epsilon_\text{i}=17.5 \text{ GeV},\quad\omega=2.35 \text{ eV (527 nm green laser)},\, \theta_\text{i}=17^o,\, \phi_\text{i}=0^o 
\label{eq:luxeparams}\end{gather}

The transition rate shows a series of Compton edges for each of the contributions from the external field (n=1,2,3...). As the intensity of the laser ($\xi$) increases, higher order contributions are more prominent but the Compton edge shifts to lower energies as a result of the larger effective rest mass ($m_\ast=m\sqrt{1+\xi^2}$) of the electron (see figure \ref{fig:hicscomp}). Clearly, it is disadvantageous in terms of producing high energy photons, to have a laser intensity much in excess of $\xi\approx 1$. \\

%\end{subfigure}\begin{subfigure}[t]{.5\textwidth}
%\end{figure}\begin{figure}
%\centerline{\includegraphics[width=0.8\textwidth]{/home/hartin/Physics_Research/myresearch/luxe/xi2_beamenergy_hics_rate_output.pdf}}
%\end{subfigure}

For large values of the intensity parameter $\xi\gg 1$, the Bessel functions have an asymptotic form with respect to Airy functions (equation \ref{eq:lcfa})\cite{NarNikRit65,ValSoa04,Nist10}. These can be used to test the validity of the locally constant field approximation (LCFA), which argues that for high intensity, the circularly polarised field looks like a constant crossed field as far as the HICS process is concerned.
 
\begin{gather}
\J_n(z_\text{h})\rightarrow \ls\mfrac{2}{n}\rs^\frac{1}{3}\Ai\lp \ls\mfrac{n}{2}\rs^{\frac{2}{3}}\!\lp 1-\mfrac{z_\text{h}^2}{n^2}\rp\rp  
\label{eq:lcfa}\end{gather}

\begin{figure}[h!]
%\centering\begin{subfigure}[t]{0.5\textwidth}
\centerline{\includegraphics[width=0.8\textwidth]{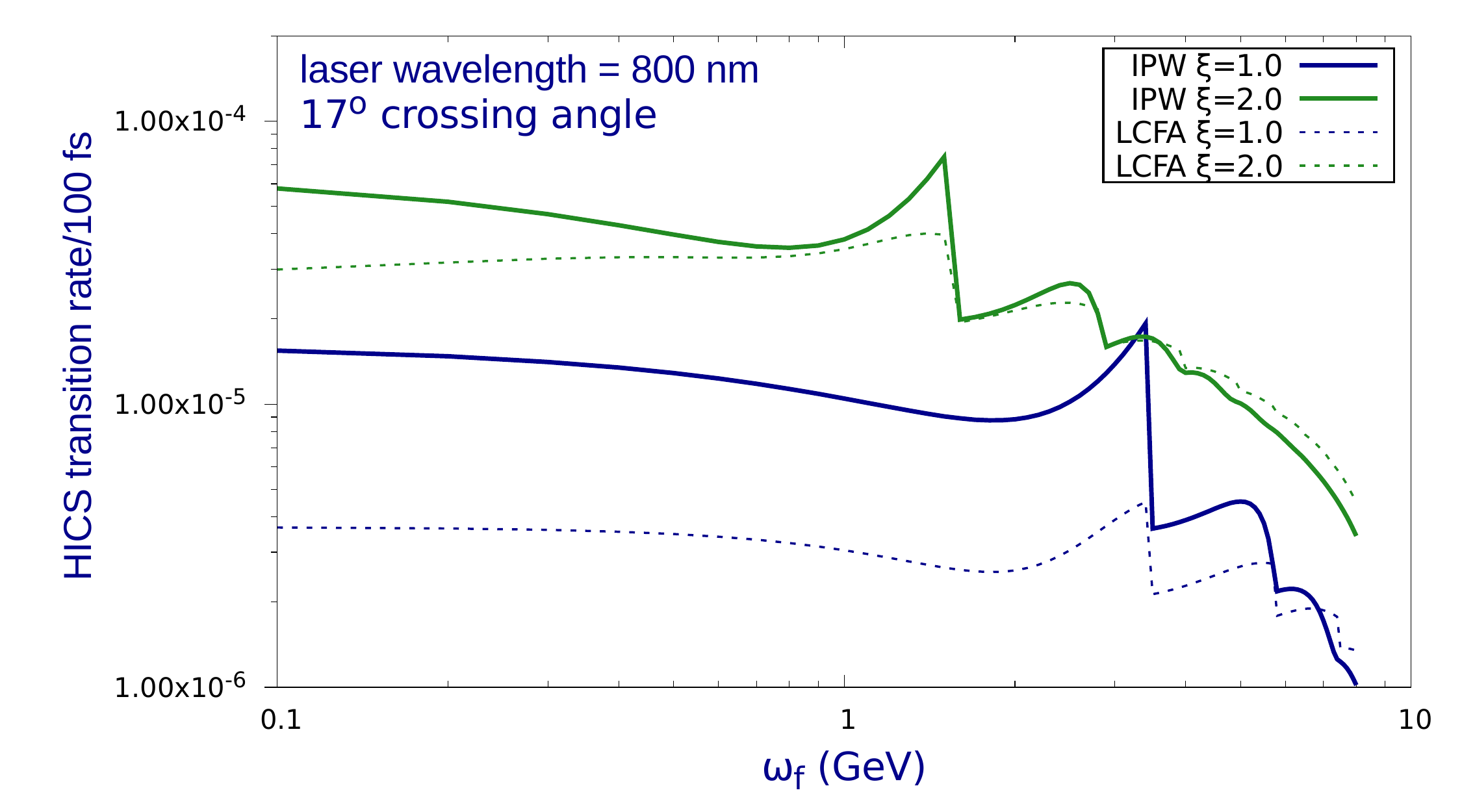}}
\caption{\bf HICS transition rate vs radiated photon energy, comparing the locally constant field approximation (LCFA) to a circularly polarised infinite plane wave (IPW).}\label{fig:hicslcfa}
\end{figure}

The LCFA can be tested numerically at the same experimental parameters specified in equation \ref{eq:luxeparams}. It appears that the LCFA does become more reasonable as $\xi$ icreases, but the validity varies also with the energy of the radiated photon, making its general applicability in a transition probability doubtful (figure \ref{fig:hicslcfa}). This underlines the advisability of using the full form of the external field rather than relying on its value local to the strong field process taking place.

\subsection{One photon pair production (OPPP)}

The other first order Furry picture process to be reviewed is the production of an electron and positron from an initial state consisting of one photon and an external field.
This process can be referred to as one photon pair production (OPPP). Due to a crossing symmetry, the OPPP matrix element can be obtained from that of the HICS with appropriate momenta swaps. The OPPP transition rate has a threshold which depends on extracting enough energy from the external field to create the pair.\\

Like the HICS process, the OPPP transition probability contains an infinite sum of amplitudes corresponding to the number of external field photons that combine with the initial photon to produce the pair. In the limit of vanishing external field intensity parameter $\xi\!\ll\! 1$, the transition probability of the OPPP process reduces to that of the Breit-Wheeler process for two photon pair production. For vanishing frequency of the external field the Toll-Wheeler result for the absorption of a photon by a constant electromagnetic field is obtained \cite{Reiss62,NikRit64a,NikRit67}. \\ 

The state of polarisation of the external electromagnetic field has a
significant effect on the OPPP process, as it does for the HICS process.
Transition probabilities for an initial photon polarised parallel and
perpendicular to a linearly polarised external electromagnetic field were
obtained by [\refcite{Reiss62,NikRit64a}]. A circularly polarised
electromagnetic field was considered by [\refcite{NarNikRit65}], and the more general
case of elliptical polarisation by [\refcite{Lyulka75}]. Spin effects were dealt
with by [\refcite{Tegnov68}]. [\refcite{Baier76}] also considered the OPPP process for an elliptically polarised external electromagnetic field using the quasi-classical operator method (section \ref{sect:collapproach}). [\refcite{Baier76}] obtained a new representation for the transition
probability in terms of Hankel functions, by considering the imaginary part
of the polarisation operator in the external field. \\

The dependence of the OPPP process on the spectral composition of the external electromagnetic field was of interest to several authors. An external electromagnetic field consisting of two co-directional linearly polarised waves of different frequencies and orthogonal planes of polarisation, yielded a transition probability of similar structure to that obtained for a monochromatic electromagnetic field\cite{Lyulka75}. \\

[\refcite{Borisov77}] considered an external field of similar form with two
circularly polarised wave components. [\refcite{Borisov77}] obtained transition probabilities
which, in the limit of vanishing frequency of one electromagnetic wave component,
reduced to those for an external field consisting of a circularly polarised field and a 
constant crossed field \cite{ZhuHer72}. In other work, several treatments of assisted pair production in non-uniform fields have appeared based on channelling phenomena in crystals \cite{Nitta04,Baier89}. A review article on the same process in heavy ion collisions and in an astrophysical setting is also available \cite{Ruffini10}. \\

The OPPP transition rate in a circularly polarised electromagnetic field has a threshold, requiring a minimum contribution from the strong field in order for there to be sufficient energy to create the pair. The threshold is expressed by a minimum value for the summation over contributions from the external field $s\geq s_0$, which depends on the 4-momenta of the initial electron $k_\text{i}$ and external field $k$, as well the external field intensity $\xi$

\begin{gather}
s> s_0\equiv \mfrac{2m^2(1+\xi^2)}{k\cd k_\text{i}}
\end{gather}

With the Volkov solution within the Furry picture, the OPPP transition probability, induced by the interaction of one quantised photon ($\omega_\text{i},\vec{k}_\text{i}$), with the external field $A^\text{e}$ and with Bessel function arguments $z_\text{v}$, is

\begin{gather}\label{eq:Woppp}
\Gamma_{\!\text{OPPP}}=\mfrac{\alpha m^2}{2\omega_\text{i}}\sum\limits_{s> s_\text{o}}^{\infty}\int_1^{v_s} \nths\mfrac{dv}{v\sqrt{v(v-1)}} \ls \J_{s}^2+\mfrac{\xi^2}{2}(2v-1)\lp\J_{s\text{+1}}^2+\J_{s\text{-1}}^2-2 \J_{s}^2\rp\rs \notag\\
z_v\equiv \mfrac{4m^2\xi\sqrt{1+\xi^2}}{k\cd k_i}\ls v\lp v_\text{s}-v\rp\rs^{1/2},\quad v_\text{s}\equiv \mfrac{(k.k_i)\,s}{2m^2(1+\xi^2)}
\end{gather}

\begin{figure}[htb]
\includegraphics[width=0.8\textwidth]{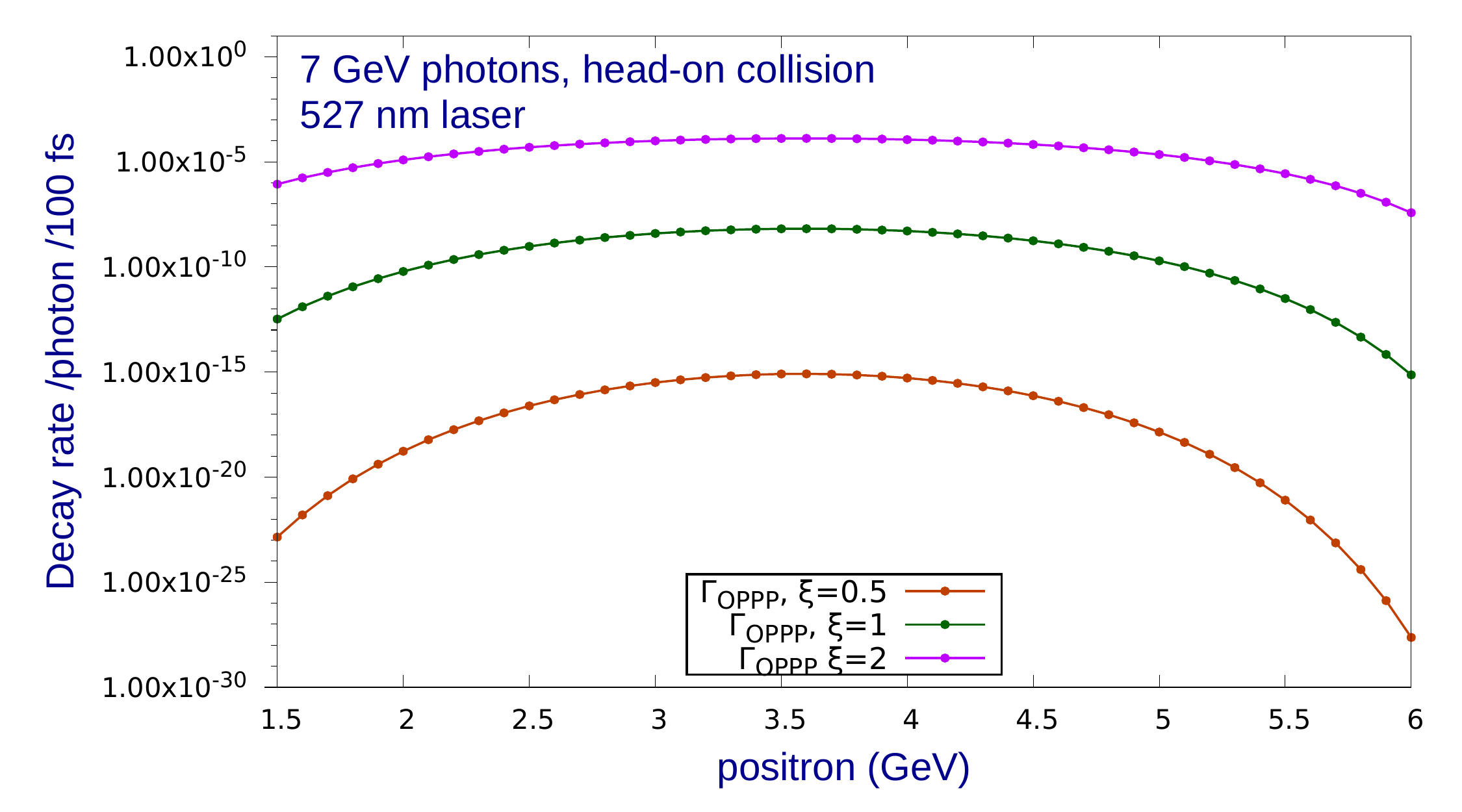}
\caption{\bf Positron spectra in OPPP pair production.}\label{fig:oppppos}\end{figure}

\begin{figure}[htb]
\includegraphics[width=0.8\textwidth]{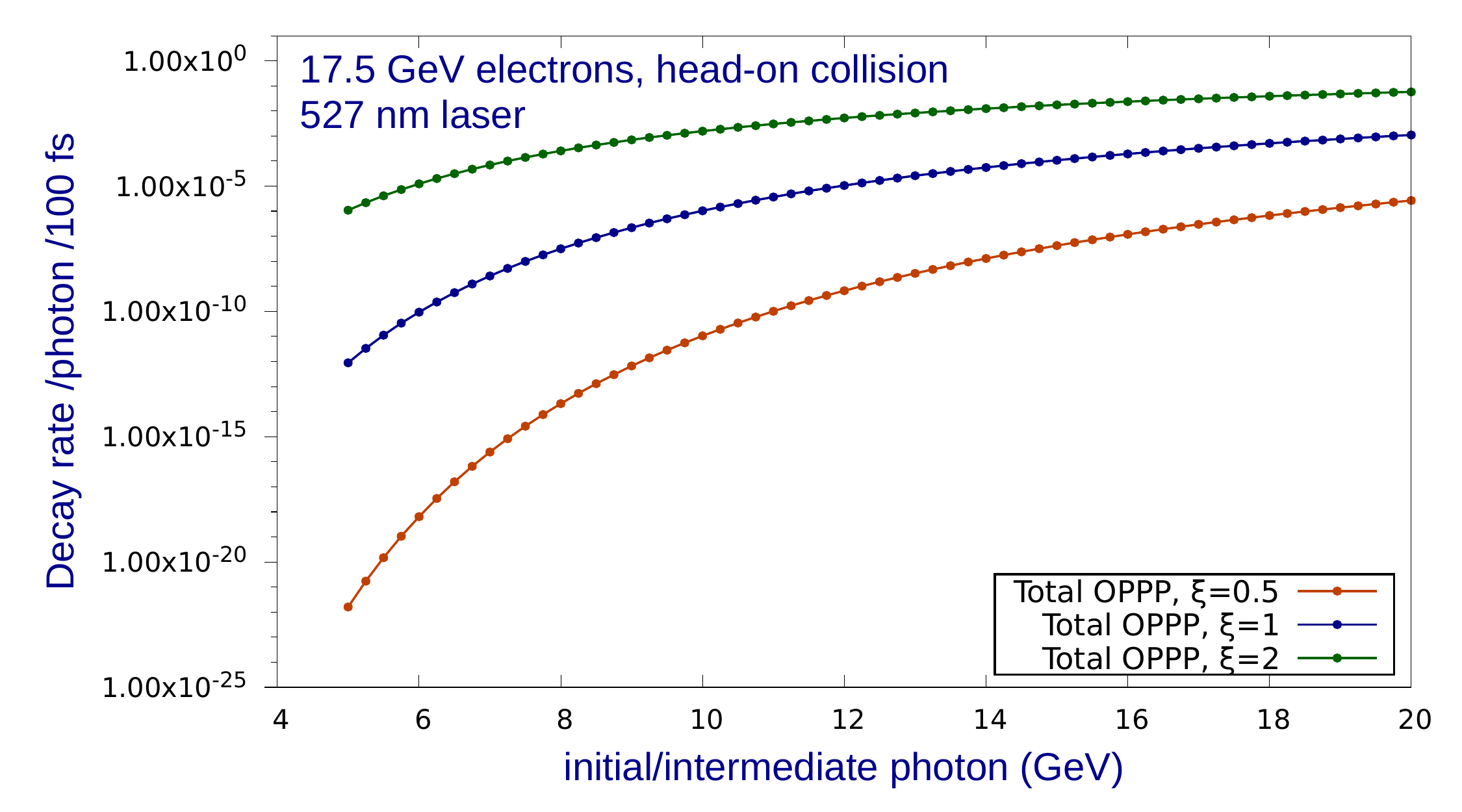}
\caption{\bf The OPPP transition rate as a function of the positron energy 7 GeV initial photons.}\label{fig:totopppt}\end{figure}

The positron spectrum produced by this process is peaked around roughly half the initial photon energy, indicating that it is more likely for the OPPP process to produce a pair which shares the initial photon energy (figure \ref{fig:oppppos}). As the strength of the laser field increases with $\xi$, there is a large improvement in the OPPP rate, as at higher intensity, more laser photons are likely to contribute their energy to producing the pair. This tendency is confirmed after integrating over allowed positron energies. The total OPPP rate versus the initial photon energy, increases from a very low value (figure \ref{fig:totopppt}). \\

\begin{figure}[htb]
\includegraphics[width=0.8\textwidth]{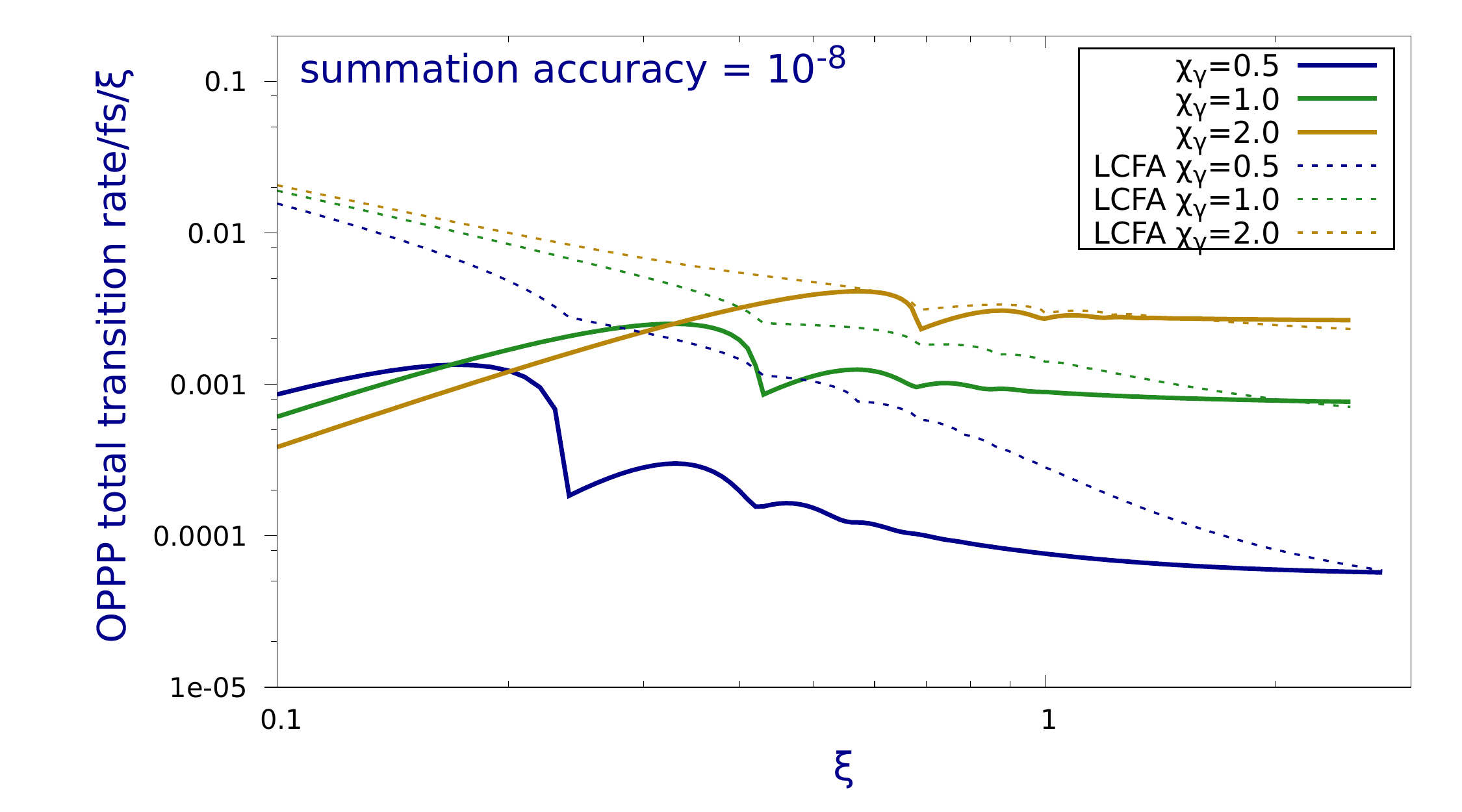}
\caption{\bf The total OPPP process as a function of external field intensity for different photon recoil parameters.}\label{fig:oppptnorm}\end{figure}

When the OPPP total rate is plotted against increasing field intensity and compared with the LCFA, the rate increases or decreases to reach a plateau depending on the value of the recoil parameter $\chi_\gamma$ (figure \ref{fig:oppptnorm}). The increase in external field intensity has a limiting effect on the process, increasing the likelihood that more external field photons will contribute to the pair production, but increasing the lepton mass and thus the energy required to produce the pair \cite{Becker91}. \\

In a real electron bunch/laser pulse collision, the HICS and OPPP process are combined, with the latter process following the former. The two processes can be linked via a real or virtual photon. The combined process with a real photon intermediary, known as the two step trident process, is considered in the next section.

\subsection{Two step trident process}\label{sect:2steptrid} 

\begin{figure}[h] 
\centerline{\includegraphics[width=0.25\textwidth]{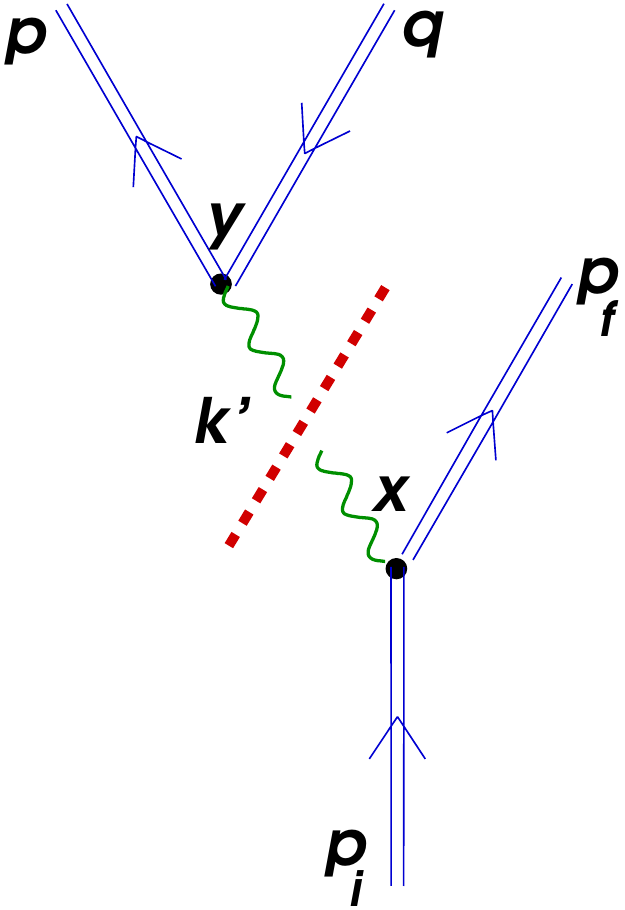}}
\caption{\bf The two step trident process with on-shell photons.}
\label{fig:2steptrident}\end{figure}

One of the signature processes of strong field experiments, involving the interaction of an intense laser field $(A^\text{e},k)$ with relativistic electrons $p_\text{i}$, is the trident process (figure \ref{fig:2steptrident}). The three final state fermions can be produced by the one step process with virtual photon (section \ref{sect:1steptrid}), or the two-step process, considered here, when the intermediate particle goes on shell \cite{Ildert11}. \\

The first attempts at experimental detection and theoretical calculation of the trident process were in the context of the E144 experiment at SLAC. The Weiszacker-Williams equivalent photon approximation was the theoretical model which was used there, and was understood to have only limited validity \cite{Bamber99}. [\refcite{Ildert11}] sketched the transition probability framework within the Furry picture, but left the details and numerical simulation to further work.

\begin{figure}[b]
\centerline{\includegraphics[width=0.8\textwidth]{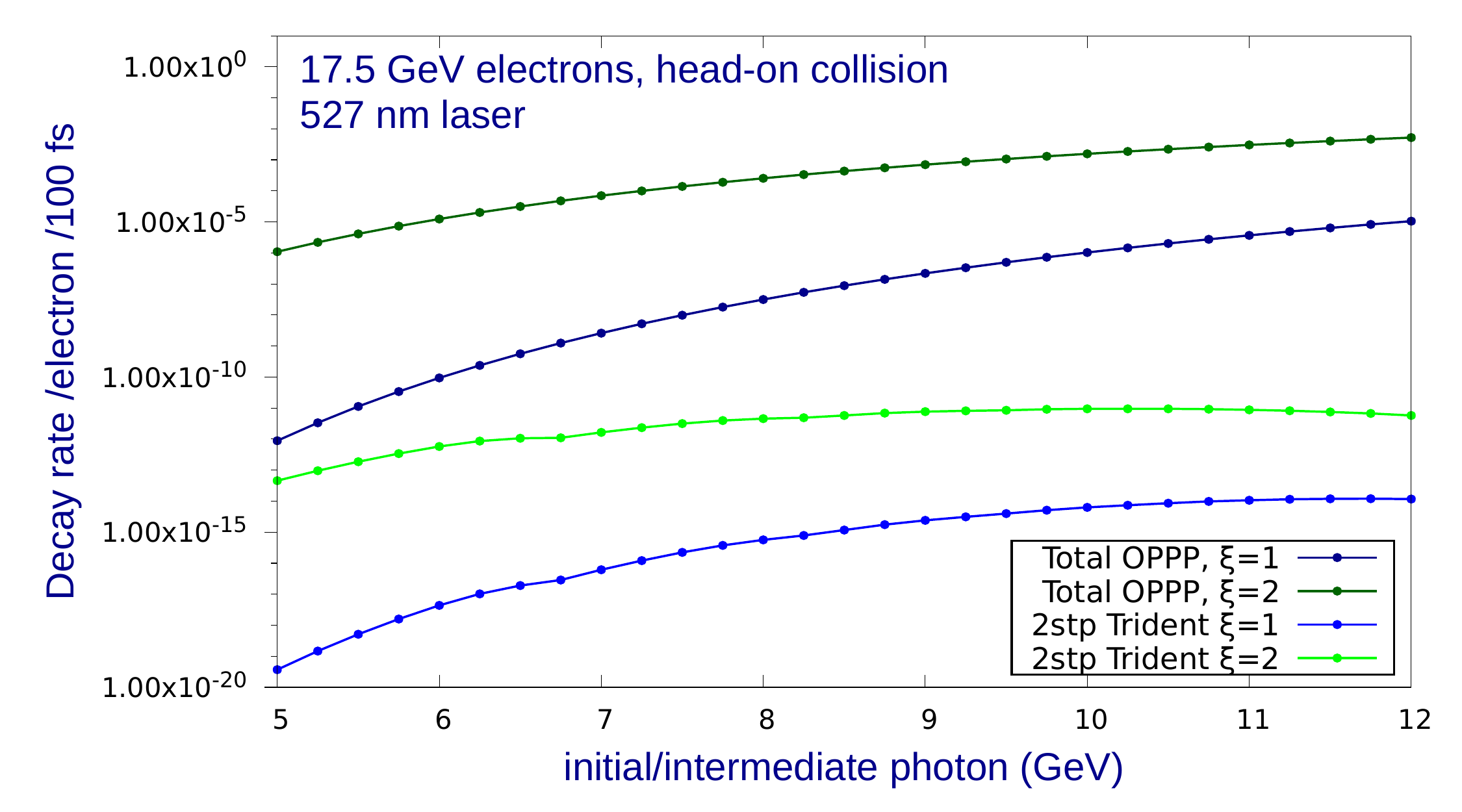}}
\caption{\bf OPPP and 2 step Trident decay rate vs initial/intermediate photon energy for an XFEL experiment.}
\label{fig:2steptridrate}\end{figure}

An effort to calculate the trident process numerically within a constant crossed field, using the rationale of the LCFA, was provided by [\refcite{King13}]. Numerical results for the trident process were presented by [\refcite{HuMulKei10}] at parameters equivalent to an imagined experiment using the european XFEL electron beam\cite{XFEL07}. The treatment of the propagator within [\refcite{HuMulKei10}] was criticised for violating gauge invariance and a more sound consideration of the trident process was provided\cite{Ildert11}. Other work to estimate the trident process numerically include a shaped laser pulse together with the LCFA and a saddle point approximation for the phase \cite{Black17}. \\

The two step trident process transition rate is simply a product of the component rates, those of the HICS and OPPP processes. Consequently, the two step trident pair production rate is some orders of magnitude lower than the rate due to the OPPP process alone (figure \ref{fig:2steptridrate}). At an actual experiment with electrons in the initial state, the two step trident process (along with the one step process) is necessarily the primary way pairs are produced. \\

\begin{table}[h!]
\tbl{\bf Strong field e$^{-}$/laser experimental parameters}{
\centering\begin{tabular}{|c|| c | c | c | c | c | c |}\hline
 Experiment & \!$\lambda(nm)$\! & \!$E_{\text{laser}}$\! & \!focus $\mu m^2$\! & \!pulse (ps)\! & \!$E_{e^-}$(GeV)\! &  $\xi$ \\ \hline\hline
   SLAC E144 & 1053 & 1 J & 50 & 1.88 & 46.6 & 0.66 \\ \hline
  XFEL/$\xi$1  & 800 & 2 J & 576 & 0.07 & 17.5 & 1 \\ \hline
  XFEL/$\xi$2  & 527 & 2 J & 100 & 0.05 & 17.5 & 2 \\ \hline
  XFEL/$\xi$3  & 800 & 2 J & 100 & 0.05 & 17.5 & 3 \\ \hline
\end{tabular}\label{tab:parsets}}\end{table}

\begin{figure}[htb]
\includegraphics[width=0.8\textwidth]{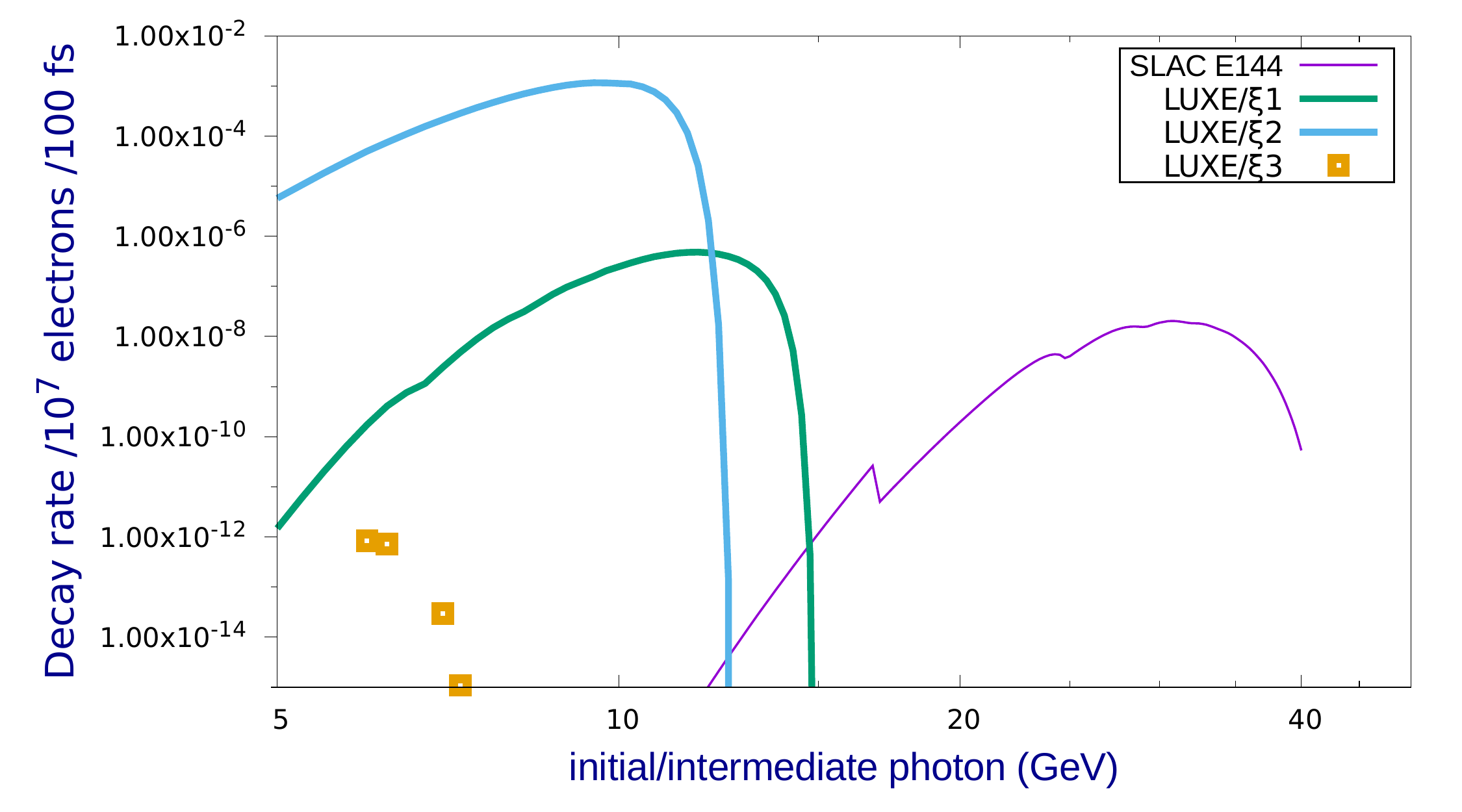}
\caption{\bf Rate of positron production via the 2 step trident process at different experiments.}\label{fig:totoppp}\end{figure}

For the beam parameters at various experiments, including the previous one carried out at SLAC E144\cite{Bamber99} and a possible future experiment, LUXE at XFEL (see table \ref{tab:parsets}), the rates of positron production can be compared. To estimate the rate for a bunch collision, a factor of $10^7$ individual interactions (that estimated at E144) is included. The rate comparison shows that there is an intermediate laser intensity $\xi\approx 2$, at which pair production is maximised. At lower intensity, the rates of both component processes are limited. At higher intensities, the HICS process produces lower energy photon for which the OPPP rate is restricted (figure \ref{fig:totoppp}). \\

A full PIC/QED monte-carlo simulation (section \ref{sect:sim}) of the trident/OPPP process and a full optimisation of parameters is required to determine ideal operating parameters for a particular experiment. \\

At this stage in the review, the standard first order strong field processes expected at dedicated electron/laser experiments have been discussed. So far, we have considered only the first order terms within the Furry picture perturbation series. The second and higher order terms have strong field propagators and have their own unique features. Some of these higher order processes are considered next.

\section{Higher order external field QED processes}\label{sect:res}

The higher order IFQFT processes provide an abundance of phenomena for study. These processes are represented by Furry picture Feynman diagrams with two or more vertices and include Compton scattering, pair production and annihilation, M\"oller scattering and the self energies. The higher order IFQFT processes require the external field photon propagator \cite{Schwinger51,Oleinik67} or the external field electron propagator which is available either in a proper time representation \cite{Oleinik68} or a Volkov representation\cite{NikRit64a,NikRit64b,Mitter75}. A 1975 review of work done on second order IFQFT processes was provided by [\refcite{Mitter75}]. \\

The possibility that second order IFQFT differential cross sections could contain resonant infinities, was recognised soon after the initial first order calculations were performed. Increasing intensities of available lasers led to the consideration of cross section terms involving contributions from two or more external field quanta. It was recognised that these contributions would lead to infinite electron propagation functions. The solution proposed was the inclusion of the electron self energy\cite{NikRit65}. \\

In the Furry picture, the mere existence of one vertex processes with a single electron or photon in the initial state demands a new physical picture. This physical picture helps to explain the unusual resonant features present in higher order processes with more than one vertex.

\subsection{The dispersive vacuum and resonant transitions}

In a non perturbative treatment of external fields, standard processes such as photon emission from electrons, can be interpreted differently. With the external field relegated to the background, the electron can be viewed as becoming unstable and spontaneously radiating. The non perturbative treatment of an external electromagnetic field, in effect, creates bound electron states and renders the vacuum a dispersive medium\cite{RafMul85,Hartin17a}. \\

The concept of dispersion in the strong field vacuum, carries over to higher order processes, when a virtual particle is exchanged between initial and final states. The virtual particle probes the strong field vacuum and sees a series of vacuum quasi energy levels. Certain final states arising from given initial states are preferred, resulting in resonant transitions - in analogy to resonant transitions between atomic energy levels \cite{Zeldovich67,Greiner85}. \\

The widths of these predicted resonances correspond to the lifetime of the electron before it decays in the dispersive vacuum/external field background \cite{Oleinik68}. The same phenomena is linked to the value of the electron self energy via the optical theorem which is equally valid in the Furry picture. The self energy is altered by the strength of the external field and is itself a Furry picture process\cite{Ritus72,BecMit76}. \\

The vacuum polarisation is expressed by the photon self energy. This involves an electron/positron loop which couples to the external field, so Furry picture resonances manifest also in the vacuum polarisation tensor. The measurement of these resonances, via an external field whose direction and strength can be varied, would give us additional information about vacuum polarization and related effects, complementary to existing schemes \cite{Heinzl06,Valle13,KingElk16}. \\
 
Resonant transitions are likewise predicted in strong field M{\o}ller scattering, where a virtual photon is exchanged\cite{Bos79a}. Both Compton scattering and two photon pair production in a strong field exhibit resonances through virtue of their virtual particle exchange \cite{Oleinik72,Hartin06}. The universality of the predicted resonant behaviour emphasises that it is a quantum vacuum related effect. \\

Some of these resonant processes, which are amenable to experimental investigation with electron/gamma/laser interactions, are now examined in more detail.

\subsection{Stimulated Compton scattering (SCS) and pair production (STPPP)}\label{sect:scs}

\begin{figure}[t]
%\centering\begin{subfigure}[t]{0.5\textwidth}
\centerline{\includegraphics[width=0.8\textwidth]{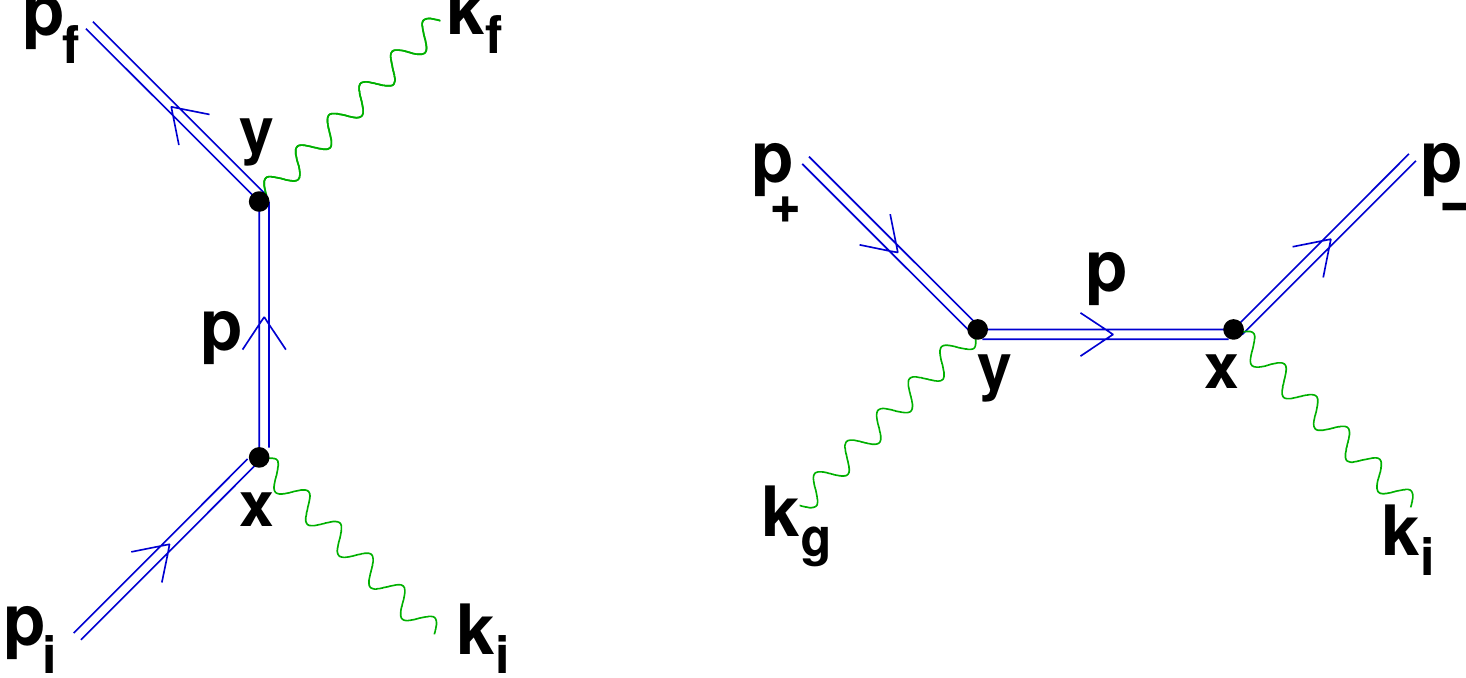}}
\caption{\bf Resonant two vertex processes in the Furry picture. Stimulated Compton scattering (SCS) and stimulated two photon pair production (STPPP) are related through a crossing symmetry.}\label{fig:2ndorder}
\end{figure}

Compton scattering in an external field, or stimulated Compton scattering (SCS, figure \ref{fig:2ndorder}) was first considered with an external field consisting of a linearly polarised electromagnetic plane wave \cite{Oleinik68}. The cross section was calculated in the non relativistic limit of small photon energy and external field intensity, for a reference frame in which the initial electron is at rest. Resonances in the cross section were present due to the poles of the electron propagator in the external field being reached for physical values of the energies involved.\\

The two photon, electron-positron pair production process in the presence of an external electromagnetic field or stimulated two photon pair production (STPPP), using Volkov solutions and without kinematic approximations, was considered by [\refcite{Hartin88,Hartin06}]. Earlier, [\refcite{KozMit87}] dealt with the process in a strong magnetic field for the case in which the energy of each of the photons is alone insufficient to produce the pair. The cross section obtained also contains resonances. \\

Resonant cross-sections occur when the energy of the incident or scattered photon is approximately the difference between two of the electron quasi-levels in the dispersive vacuum \cite{Zeldovich67}. The cross section resonance widths were calculated by inserting the external field electron self energy into the electron propagator. The resonant cross section exceeded the non resonant cross section by several orders of magnitude \cite{Oleinik67}. \\

[\refcite{AkhMer85}] considered the SCS cross section in a linearly polarised external electromagnetic  field and wrote down the SCS matrix element for a circularly polarised external field. This calculation was performed for the special case where the momentum of the incoming photon is parallel to the photon momentum associated with the external field. [\refcite{AkhMer85}] avoided the resonant infinities by considering a range of photon energies for which resonance did not occur. \\

The full SCS cross-section, without kinematic approximations was calculated for a circularly polarised external field and extensive numerical scans were performed \cite{Hartin06}. It was realised that the resonances could be exploited in order to produce a source of energy up shifted photons \cite{Hartin17b}. Resonances in these processes have also been considered in a series of papers by [\refcite{Roshchup96}]. When allowance is made for the pulse length within a real laser interaction, these resonances are generally broadened by the extent to which the IPW approximation differs from the pulsed treatment \cite{Roshchup12,Seipt12}. \\

Here, the main analytic features of the SCS cross-section are described with an external field consisting of a circularly polarised plane wave, $A^\text{e}(k\cd x)$. The initial photons $k_\text{i}$ should be provided by a tunable source, so that the energy is variable. For the STPPP process, a source of high energy photons is required in order to reach the threshold of the pair production. \\

The main feature of interest is the structure of the propagator $G^\text{FP}_\text{yx}$ and its subsequent effect on the transition probability. The analysis begins with writing down the matrix element with the aid of the Furry picture Feynman diagrams (figure \ref{fig:2ndorder}). The notation has $\Psi^{\text{FP}\pm}_\text{fry}$ representing the positive and negative energy $(\pm)$ Volkov solutions for the state with momentum $p_\text{f}$ and spin $r$ at vertex $y$. $A_\text{f,i,g}$ are the quantised photons interacting with the Volkov fermions and the laser field is $A^\text{e}$,

\begin{gather}
M^\text{SCS}_\text{f\,i}= \int \text{d}x\, \text{d}y\; \bar \Psi^{\text{FP}+}_\text{fry}\,\bar A_\text{fy}\,G^\text{FP}_\text{yx}\,A_\text{ix}\,\Psi^{\text{FP}+}_\text{isx}\notag\\
M^\text{STPPP}_\text{f\,i}= \int \text{d}x\, \text{d}y\; \bar \Psi^{\text{FP}-}_\text{fry}\,\bar A_\text{gy}\,G^\text{FP}_\text{yx}\,A_\text{ix}\,\Psi^{\text{FP}+}_\text{isx}\\
G^\text{FP}_\text{yx}=\medint\int\mfrac{\text{d}p}{(2\pi)^4}E_\text{py}\;\mfrac{\st{p}+m}{p^2-m^2+i\epsilon}\;\bar E_\text{px},\quad  \Psi^{\text{FP}\pm}_\text{fry}= n_\text{p}\,E^\pm_\text{fy}\; u^\pm_{\text{fr}}\;e^{\mp i p_\text{f}\cdot y }\notag
\end{gather}

As was done for the one vertex Furry picture processes, the next step is to consider the dressed vertices and Fourier transform them in order to extract the modes,

\begin{gather}\label{eq:FTVolkov}
E_\text{fy}\gamma_\mu \bar E_\text{py}= \sum^\infty_{n_\text{y}=-\infty}\medint\int^{\pi}_{-\pi} \mfrac{d\phi}{2\pi} \; E_{\text{f}\phi}\gamma_\mu \bar E_{\text{p}\phi}\;\;e^{- i (p_\text{f}-p+n_\text{y}k)\cdot y }\;
\end{gather} 

In a multi-vertex Furry picture diagram, there will appear more than one sum over extracted modes, in this case two sums labelled $n_\text{x},n_\text{y}$. In the usual treatment, the sums are shifted $n\equiv n_\text{x}+n_\text{y},\,\,l\equiv n_\text{x}-n_\text{y}$ so that only one contribution appears in the overall momentum conservation. \\

With these steps, the transition probability of the two vertex SCS and STPPP processes can be expressed as an overall sum of contributions of photons from the strong field $nk$, with internal contributions $lk$. The action of the external field also induces a momentum/mass shift in fermions $p_\text{i}\rightarrow q_\text{i}$ which is dependent on the relative direction of motion and the intensity of the external field. Thus, the conservation of momentum for the SCS and STPPP processes can be written,

\begin{gather} \label{eq:consmom}
q_i+k_i+nk\rightarrow q_f+k_f \quad \text{SCS} \notag\\
k_g+k_i+nk\rightarrow q_{+}+q_{-} \quad \text{STPPP} \\
 n\in\mathbb{Z}\quad m^2=1+\xi^2,\quad q_i=p_i+\mfrac{\xi^2\,m^2}{2k\cd p_i}k,\quad \xi\equiv\mfrac{e|\vec{A}^\text{e}|}{m} \notag
\end{gather}

The conservation of energy-momentum leads to momentum flow into and out of the propagator. When this momentum flow reaches the mass shell, the propagator denominator goes to zero, blowing up the transition probability. In reality, there is an imaginary mass correction to be made to the denominator corresponding to the interaction of the virtual particle with it's own field (section \ref{sect:selfene}). With the mass correction, the propagator pole is regularised and a resonance appears instead of an uncontrolled infinity. \\

For the Furry picture though, there are a number of pole conditions corresponding to the sum over modes extracted from the dressed vertices. These in turn lead to a series of resonance conditions which can be described as occurring when the propagator momentum flow matches the energy difference between two Zeldovich quasi-energy levels, which are induced in the vacuum by the intense background field \cite{Zeldovich67}. \\

There are two resonant conditions both for the SCS process and STPPP processes, corresponding to the direct and exchange channels, and containing internal contributions from the external field $lk$,

\begin{gather}\label{eq:rescond} 
(q_\text{i}+k_\text{i}+lk)^2=m^2(1+\xi^2)\quad  \quad \text{SCS direct channel} \notag\\
(q_\text{i}-k_\text{f}+lk)^2=m^2(1+\xi^2)\quad \text{SCS exchange channel}  
\end{gather}

Without the external field, the mass shell condition can only be obtained for soft photons (the infra red divergence). The IR divergence is partly dealt with by the experimental constraint of finite energy detector resolution (infinitesimally low energy photons can't be detected). In contrast, the SCS and STPPP mass shell conditions can be achieved in multiple circumstances, for experimentally achievable kinematic parameters. With use of the resonance conditions, and with the conservation of energy-momentum, resonance requirements on the initial state photon and the resultant signature of the final states, can be obtained.\\

This resonant feature, of strong field particle processes with a propagator that can go on shell, has been noted before \cite{Oleinik67,Oleinik68,Bos79a,Bos79b,Roshchup96}. In order to perform experiments, it is necessary to do a full calculation of the SCS and STPPP transition probabilities. This is quite a challenging mathematical task due to the relative complexity of the Volkov spinor and phases. However, recent steps have been taken, by use of Fierz transformations of Volkov spinors, that simplify the analytic expressions \cite{Hartin16}. \\

One may speculate on the physical meaning of the quasi energy levels seemingly present in higher order Furry picture processes. An analysis of the modes extracted from the vertex dressed by Volkov solutions are that these constitute Floquet-Volkov states\cite{Seipt12}. One can surmise that there is a physical basis to the Floquet-Volkov states, just as Floquet-Bloch states are based on periodic structures in a solid \cite{Mahmood16}. The Volkov states form in the quantum vacuum, which consists in part, of charged particles which form dipoles to screen real, bare charges. With this analogy, the periodic, external electromagnetic field may induce structures among virtual vacuum dipoles, which in turn lead to Volkov states and the Zeldovich quasi-energy levels. \\

\begin{figure}[htb] 
\centerline{\includegraphics[width=0.8\textwidth]{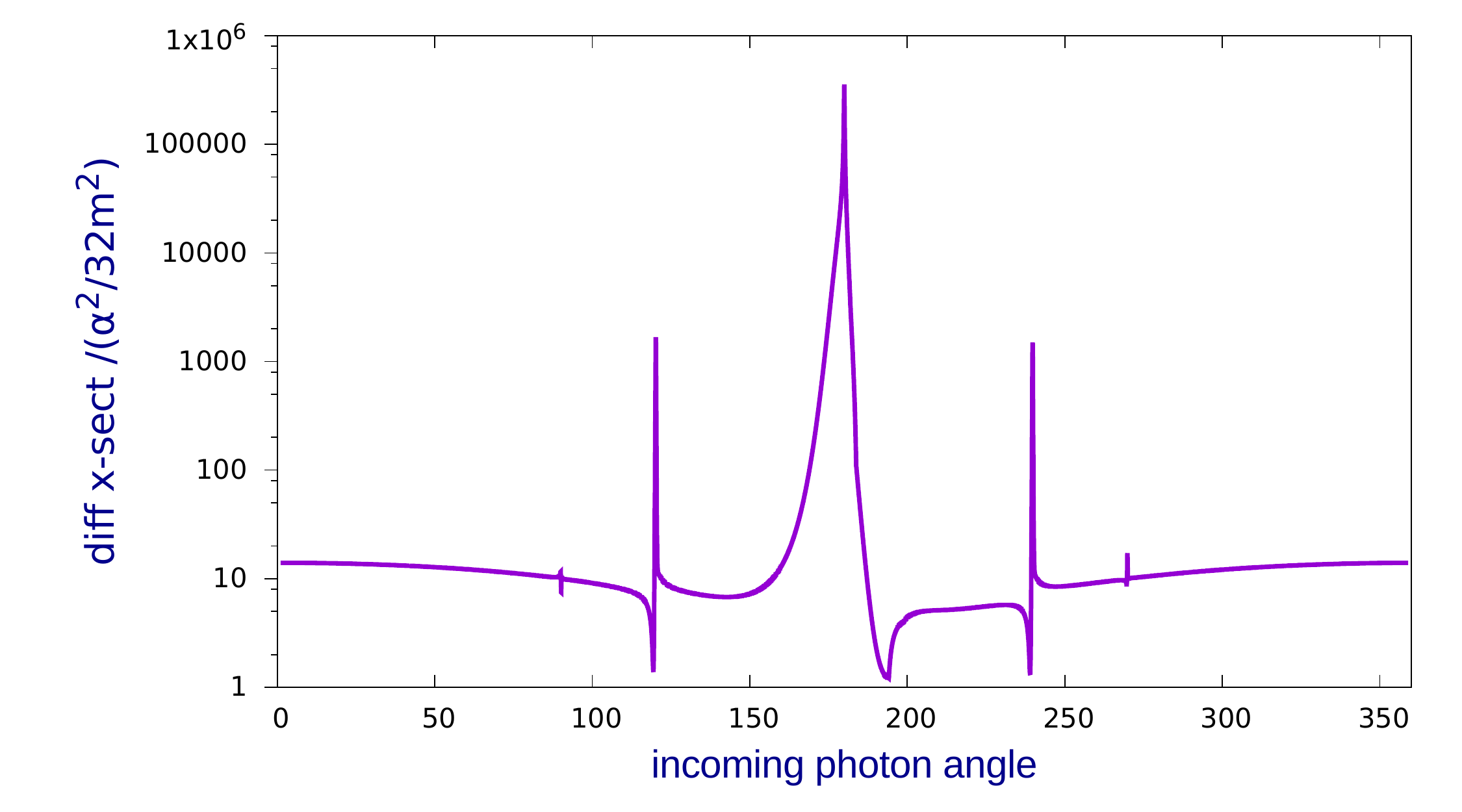}}
\caption{\bf SCS resonances for 80 MeV electrons, 4 eV initial (probe) photons, $\xi$=1,1 eV intense laser. The probe photon angle is defined by the direction of propagation of the intense laser}
\label{fig:scsresscan}\end{figure} 

In any event, the full calculation of the SCS transition probability, including the regularisation of the propagator poles can be carried out, and particular numerical studies that display resonance peaks can be performed. Such a study for an experiment in which electrons of energy 80 MeV collide head on with an intense optical laser ($\omega$=1 eV, $\xi$=1). Scans of 4 eV probe photons about the angle of incidence with respect to the laser propagation direction, traverse regions where resonance conditions are satisfied (figure \ref{fig:scsresscan}). \\

There is a broad primary resonance at $\theta_\text{i}=180^o$ (i.e. probe photons counter propagating to the intense laser), with an enhanced cross-section orders of magnitude above the baseline, corresponding to a minimum contribution (l=1) from external field photons. Reduced, off axis, side band resonances correspond to $\lv l\rv> 1$ contributions. Each resonance peak is not symmetrical about its maximum point because in general the resonance width is dependent on the kinematics as well. \\

Before going on to discuss how the propagator poles are regularised, one more higher order Furry picture process is examined.

\subsection{One step trident process}\label{sect:1steptrid} 

The one-step trident process has attracted almost as much recent attention as the two step trident process \cite{HuMulKei10,Ildert11,King13,KingFed18,DinTor18}. The reason for this interest are the planned strong field electron/laser experiments in which the one step trident process will appear as a key process. Additionally, there is a need to go beyond previous analyses that approximated the trident cross section with Weizs\"acker-Williams equivalent photons. \\

\begin{figure}[htb] 
\centerline{\includegraphics[width=0.3\textwidth]{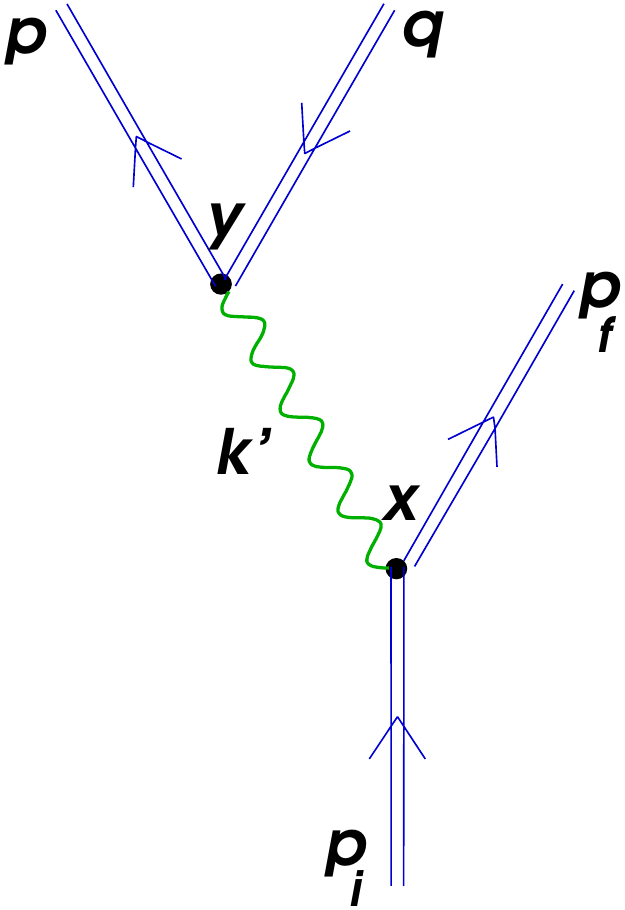}}
\caption{\bf The one step trident process.}
\label{fig:1steptrident}\end{figure} 

The strong field physics applicable to the one step trident process, compared to the two step, is in the photon propagator and its resonance conditions (see figure \ref{fig:1steptrident}). At first sight it may appear strange that the photon, which does not couple to the external field, is implicated in resonant transitions. However, these transitions are determined by the dressed vertices and the momentum flow. Moreover, the virtual photon sees a vacuum that is polarised by the intense laser field and couples to it through its self energy via a Furry picture virtual loop. \\

The one step trident process has two channels, involving the swap of the final electron-positron momenta ($p\leftrightarrow q$). Since the momentum flow remains the same however, there is only one series of resonance conditions expressed by,

\begin{gather}
\ls p+q+m^2\xi^2k\lp\mfrac{1}{k\cd p}+\mfrac{1}{k\cd q}\rp+lk\rs^2=0
\end{gather}

The one step trident transition probability calculation is straightforward, though lengthy. The analytic work required to calculate the trace terms could be simplified by application of the Fierz method for Volkov spinors\cite{Hartin16}, so that no kinematic approximations need be applied. \\

\begin{figure}[htb]
\centerline{\includegraphics[width=0.4\textwidth]{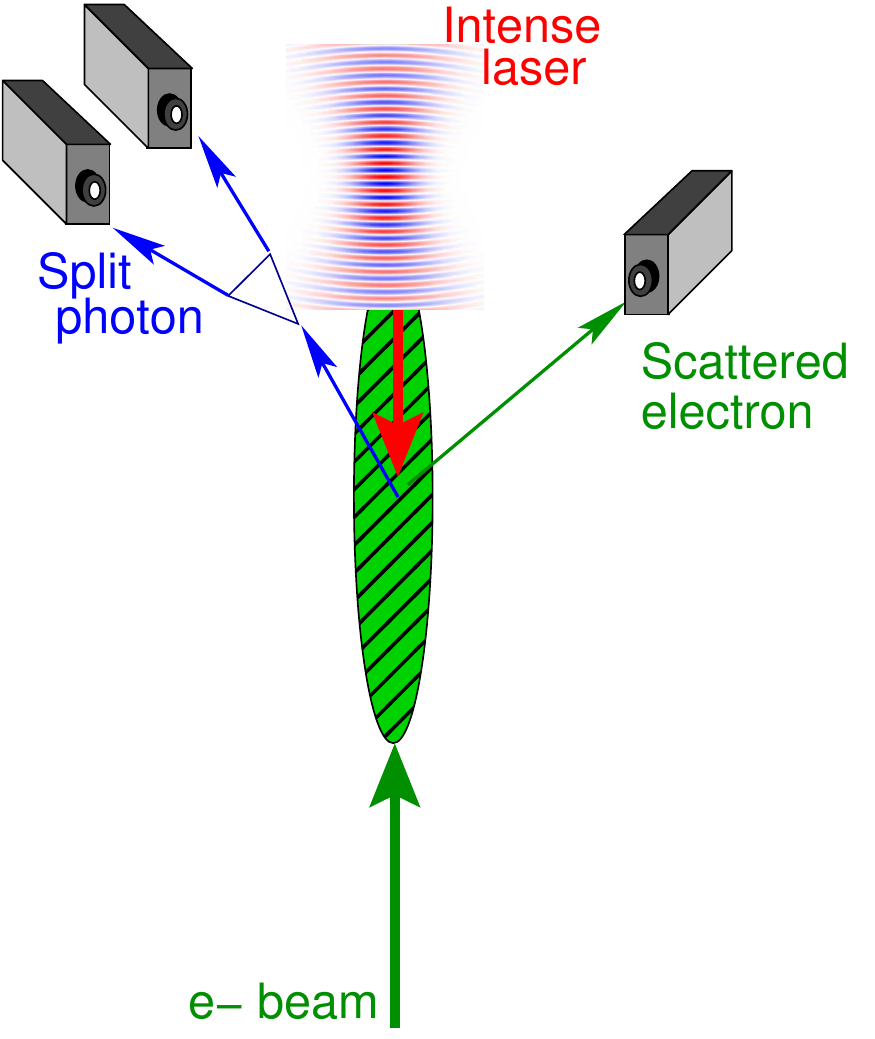}}
\caption{\bf Experimental set up for photon splitting via the one step trident process.}\label{fig:photsplit}
\end{figure}

The experimental appearance of these Furry picture resonances, in any of the strong field processes reviewed here, would be a dramatic confirmation of the higher order theoretical predictions of the Furry picture. These resonances, once found, could also be exploited for searches for additional new physics. For instance, the trident process can also produce a pair via a three vertex photon splitting loop (figure \ref{fig:photsplit}). This is a low probability process, however its transition probability is boosted at resonance by potentially orders of magnitude. Detailed studies of these and other rare transitions are under way. \\

In order to regulate the propagator poles in the tree level, higher order Furry picture processes there is further theoretical work with Furry picture loops to be done.

\subsection{Furry picture self energies}\label{sect:selfene}

\begin{figure}[htb]
\centerline{\includegraphics[width=0.5\textwidth]{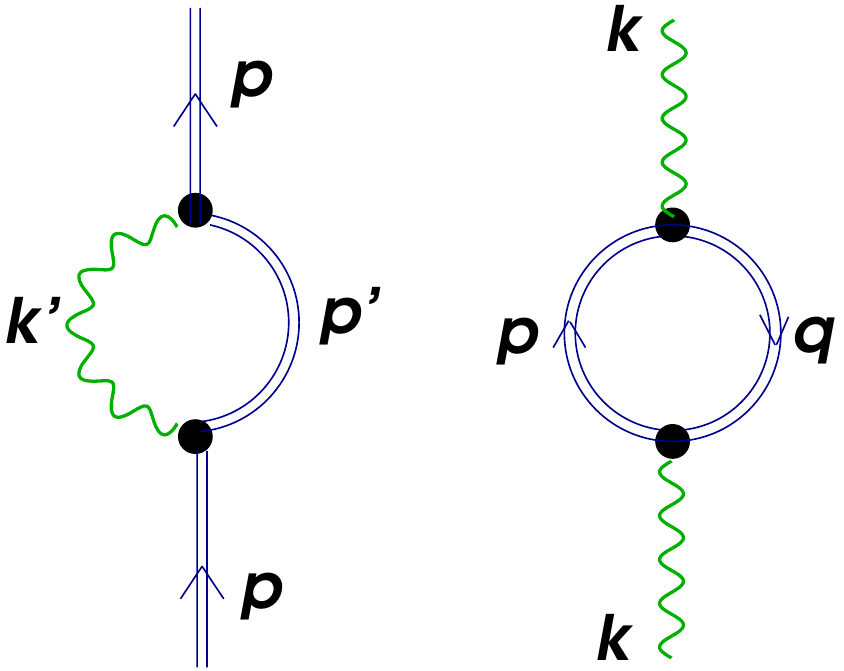}}
\caption{\bf Furry picture self energy Feynman diagrams. These are related to the Furry picture one vertex HICS and OPPP processes through the optical theorem.}
\label{fig:selfene}\end{figure}

The consistent treatment of the resonant cross sections of higher order IFQFT processes, require the calculation of the electron and photon self energies in the presence of an external field (figure \ref{fig:selfene}), and their insertion into strong field propagators. \\

[\refcite{Ritus70}] was one of the first to consider the electron and photon self energy 
processes in an external electromagnetic field using Volkov solutions. The vehicle for the calculation was the mass operator, which was considered for a constant crossed field \cite{Ritus72}, circularly polarised field\cite{BecMit76} and a combination of constant and elliptically polarised fields\cite{Klimenko90a,Klimenko90b}. Via the mass operator, the mass correction to the probability for one photon emission, and the mass correction to the anomalous magnetic moment of the electron were found \cite{MorRit75}. In important work on a Furry picture effective coupling constant, asymptotic expressions were obtained to calculate the corrected electron propagator to third order, which yielded an estimation of the lower bound for an IFQFT expansion parameter\cite{Narozhnyi79}. \\

The Furry picture photon energy is obtained via the vacuum polarisation operator. [\refcite{Narozhnyi69}] calculated this in a constant crossed field and demonstrated propagation modes with different group and phase velocities\cite{KleNig64,Bialynicka70}. Photon elastic scattering was studied with the vacuum polarisation operator\cite{Sannikov67,Sannikov95}, as was pair production\cite{MorNar77}. Treatments that exploited gauge, charge and relativistic invariance proved insightful\cite{BatSha71,BecMit75}. Light cone coordinates enable a simplification\cite{NevRoh71}, whereas a proper time representation for propagators lead to cumbersome expressions\cite{AffKru87}. Other techniques used to perform calculations, were renormalisation group methods \cite{ColWei73,Kryuchkov80} and a method adapted from string theory\cite{Schubert01}. The vacuum polarisation operator was obtained in other fields including a Coulomb field \cite{Yakovlev67}, a magnetic field\cite{Shabad75} and a circularly polarised electromagnetic field\cite{BecMit75}.\\

The usual procedure for self energy corrections is to insert them into propagator denominators via a geometric sum\cite{PesSch95}. There are, however, difficulties with this procedure for the Furry picture, in that the virtual particle is coupled not only to its own field, but to the external field as well. The Furry picture self energy does not have a simple dependence on the propagator momentum, and the usual interpretation and absorption of divergent parts is complicated \cite{BecMit76}. \\

Another path forward is by considering the physical picture of a dispersive vacuum that the Furry picture suggests. Both the photon and electron are unstable in such a vacuum and will decay to a pair or lose energy through photon radiation, respectively. Then, one calculates a lifetime for the virtual particle and includes it via the usual procedure for unstable states\cite{PesSch95}. \\

Vacuum dispersion implies that intermediate virtual states contain a complex mass shift $M^\text{FP}_\text{p}$ and associated resonance. To first order, the mass shift includes one loop Furry picture diagrams. Propagator poles will be moved by the real part of the mass shift and the imaginary part will constitute the resonance width. By use of the LSZ formalism, the coupling constant will also be adjusted from its bare value to its observable value ($e\equiv \sqrt{Z}e_0$) \cite{PesSch95}. Taking the Furry picture electron propagator $G^\text{FP}_\text{yx}$ and self energy $\Sigma^\text{FP}_\text{p}$ as an example,

\begin{gather}
G^\text{FP}_\text{yx}\!=\!\medint\int\mfrac{\text{d}p}{(2\pi)^4}E_\text{py}\,\mfrac{ie_0^2}{p^2-m_0^2+i\epsilon}\,\bar E_\text{px}\rightarrow \medint\int\mfrac{\text{d}p}{(2\pi)^4}E_\text{py}\,\mfrac{iZe_0^2}{p^2-m_0^2+\mathfrak{R}M^\text{FP}_\text{p}+i\mathfrak{I}M^\text{FP}+i\epsilon}\,\bar E_\text{px} \notag\\[4pt]
M^\text{FP}_\text{p}\equiv \Sigma^\text{FP}_\text{p}+... \, , \quad 
\Sigma^\text{FP}_\text{p}=ie^2\sum_\text{spins}\int \text{d}u\,\text{d}v\,\bar E_\text{pu\,}\gamma^\mu\, G^\text{FP}_\text{uv} \,D^\text{FP}_\text{vu} \,\gamma_\mu \, E_\text{pv}
%\mathfrak{I}\Sigma^\text{FP}_\text{p}\equiv \Gamma=2 W^\text{HICS} \notag
\end{gather}

The LSZ formula relates the one particle irreducible (1PI) diagrams to the forward scattering. The largest leading term in the 1PI sum, the Furry picture self energy, $\Sigma^\text{FP}_\text{p}$ (figure \ref{fig:selfene}) is related to the HICS radiation process via the optical theorem. To leading order then, the resonance width $\Gamma$ is,

\begin{gather}
\mathfrak{I}M^\text{FP}_\text{p}|_\text{leading order} =\mathfrak{I}\Sigma^\text{FP}_\text{p}=2 W^\text{HICS}\equiv\Gamma
\label{eq:gammafull}\end{gather}

The imaginary and real parts of the mass correction $M^\text{FP}_\text{p}$ are related to each other through a dispersion relation, and both are related to the observable coupling constant. A precise experimental determination of Furry picture, resonance poles and widths, would provide sensitive new tests, both for QED and for the more general quantum field theories in background potentials. \\

The procedure for inclusion of self energies in the Furry picture seems straightforward, however there are still open questions. The first question regards the renormalisation of the interacting mass and charge, which remains necessary even if the interaction with the external field is exact. The procedure requires regularisation and absorption of UV divergences into the observable mass and charge. The renormalisation scheme must take into account the tensor structure of the Furry picture self energy. There are terms coupling the spin to the self field as well as the external field \cite{BecMit76,Ritus79}. One possibility is that the mass renormalisation contains spin dependent as well as external field dependent terms.\\

The second question regards the UV divergence itself. Physically, it is understood as arising from beyond standard model (BSM) physics, that modifies the theory at small distance scales. In the Furry picture, with a background field which modifies the interactions at all scales, it is possible that the UV divergences are automatically regularised. \\

Another theoretical challenge regards the running of the coupling constant as the external field strength grows towards the Schwinger critical value. Since the external field couples with the self energy to the electron, the field strength plays a part in the coupling constant. If the external field strength is large enough, the sum of 1PI diagrams appears to no longer converge \cite{Narozhnyi79}. \\

For inclusion of Furry picture self energies, a LSZ-type schema consistent with IFQFT must be utilised. The LSZ theory is not easily extended to bound states. The background field means that the interaction particles do not strictly form asymptotically free states. Techniques from QFT in curved space-times which seem to have dealt with this problem \cite{Wald10} can possibly be of use. \\

The background field, which distinguishes space-time intervals, also violates Poincar\'{e} invariance. So the strong field propagator depends on separate space-time points and not on the difference between them. A reformulation of the Furry interaction picture via an algebraic approach in which quantum fields are local and co-variant, is suggested. To re-establish Poincar\'{e} invariance, a short distance operator could be implemented \cite{Wald10}. \\

These open theoretical questions are still under consideration. In any case, one can determine experimentally where in parameter space resonant transitions are likely to occur. Experimental designs and requirements can already be explored with the aim of shedding light on theoretical predictions and challenges.

\section{Experimental schemas for resonance detection}\label{sect:exp}

Intense laser/electron beam interactions have already been performed successfully, in order to produce strong field signatures of first order Furry picture processes\cite{Bamber99}. The HICS, strong field radiation process (section \ref{sect:hics}) was induced in the head-on collision of 46.6 GeV electrons and an optical laser focussed to an intensity of order $10^{18} \text{ Wcm}^{-2}$. \\

\begin{figure}[h]
\centerline{\includegraphics[width=0.8\textwidth]{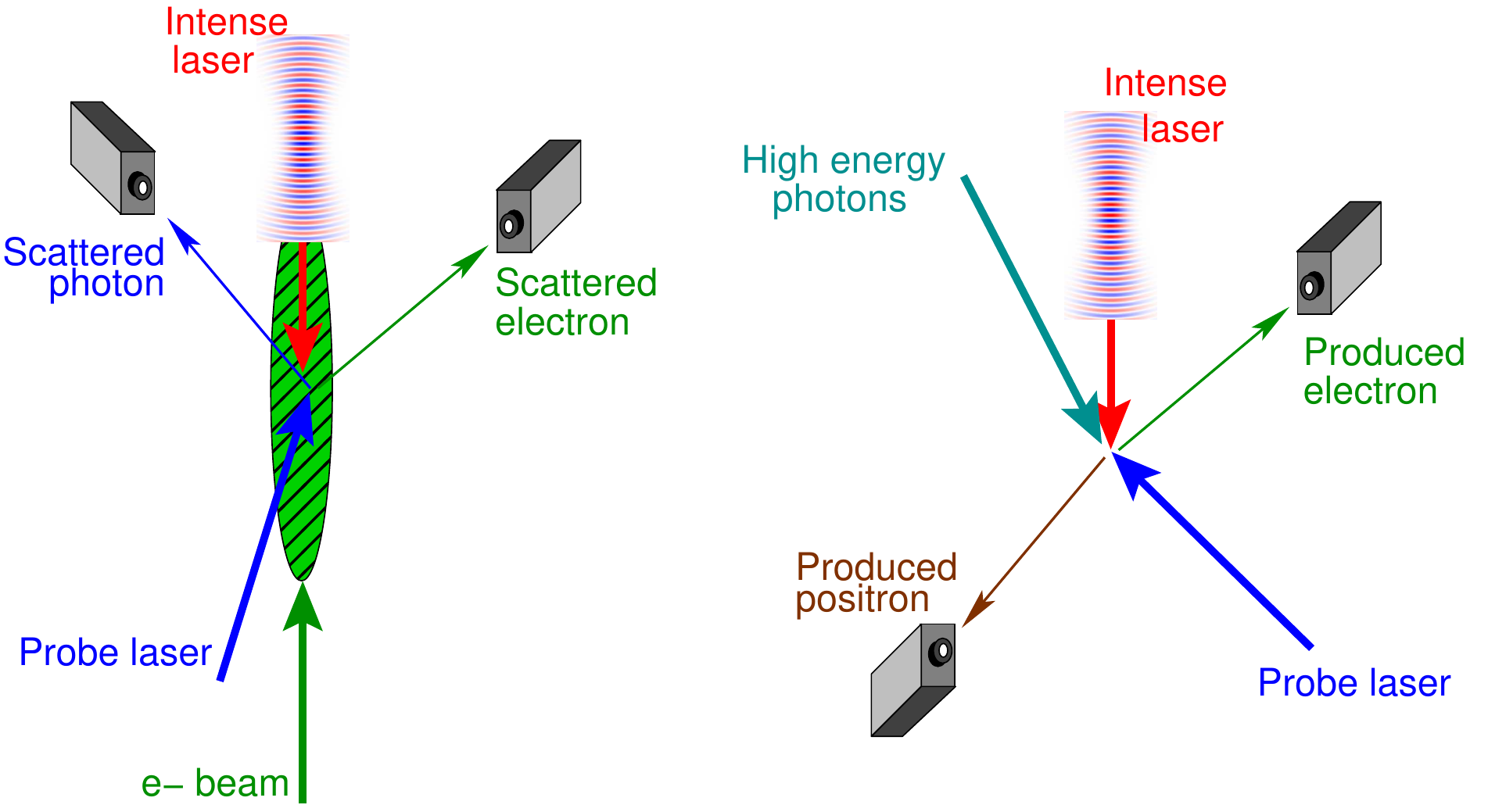}}
\caption{\bf Experimental set up for the SCS and STPPP processes.}\label{fig:scsexptconfig}\vspace{0.1cm}
\end{figure}

In order to experimentally produce the resonances of the SCS process, a similar, near head-on collision can be set up, preferably with primary inverse Compton scattered photons directed away from the final state region of interest. Resonances can then be scanned over by introducing a probe laser whose direction and/or energy can be varied. Scattered particles are captured with suitable detectors which can scan the range of final state angles and energies (figure \ref{fig:scsexptconfig}). \\

For the STPPP process, a source of high energy photons sufficient to reach the threshold of pair production, should interact with the strong laser at some small incident angle to inhibit one photon pair production processes. Scans can once again be made over probe photon energies and angles to locate resonant STPPP pair production. \\

The analysis continues here only for the SCS process. Similar relations and features will pertain to the STPPP process, given that production threshold is reached\cite{Hartin06}. \\

In terms of experimentally available parameters, one can assume that a table top, optical laser focussed to $10^{19} \text{ W cm}^{-2}$ is readily available. This corresponds to a strong field intensity parameter of $\xi \gtrsim 1$, though intensity can be tuned lower with longer pulse lengths, or less stringent focussing. It is desirable also to adjust laser parameters so that a given intensity is achieved in as long a laser pulse as possible, in order to enhance resonance peaks.\\

\begin{figure}
\centering
\includegraphics[width=0.3\textwidth]{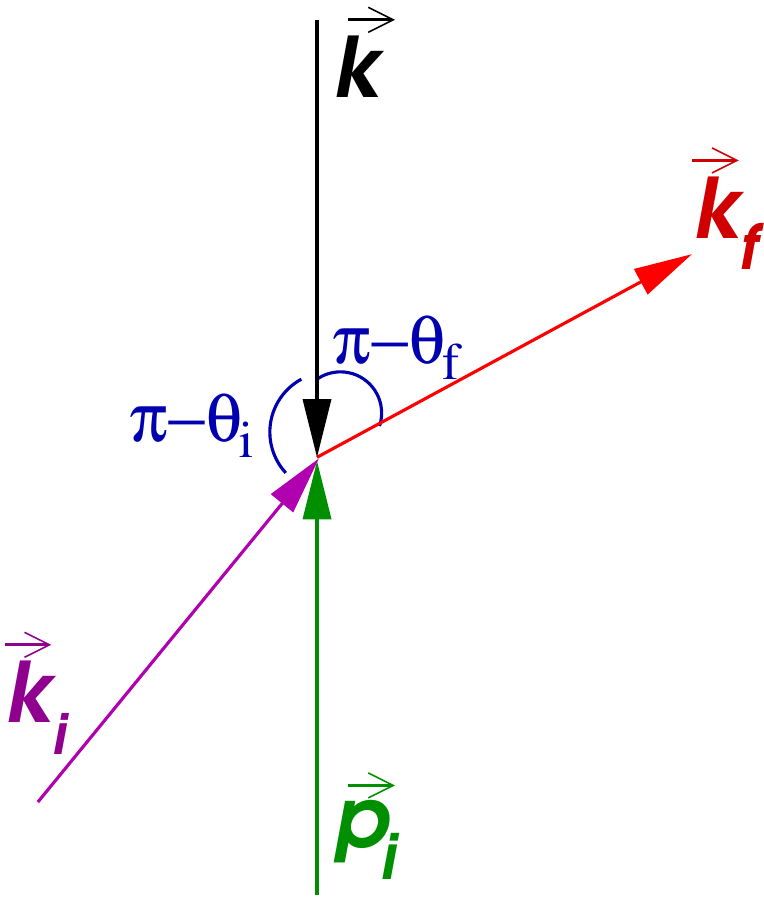}\caption{\bf probe and radiated photon angles in the SCS process.}
\label{fig:scsang}\end{figure}

To induce SCS resonances, electron beams must be brought into coincidence with the intense laser and a probe laser. Scattered photons should be detectable over a range of scattering angles. The probe laser should be a tunable optical laser, with a variable energy $\omega_\text{i}$ and angle of incidence to the external field propagation direction, $\theta_\text{i}$ (figure \ref{fig:scsang}). \\

For relativistic electrons ($\gamma,\beta$) colliding head-on with a strong laser of intensity $\xi$ and energy $\omega$, the condition for the lth level resonance can be obtained ($l=1$ is the primary resonance, $l=2$ the secondary, etc). The direct channel gives a resonant probe laser incident angle, and the exchange channel gives a resonant radiated photon angle,

\begin{figure}
\centering
\includegraphics[width=0.8\textwidth]{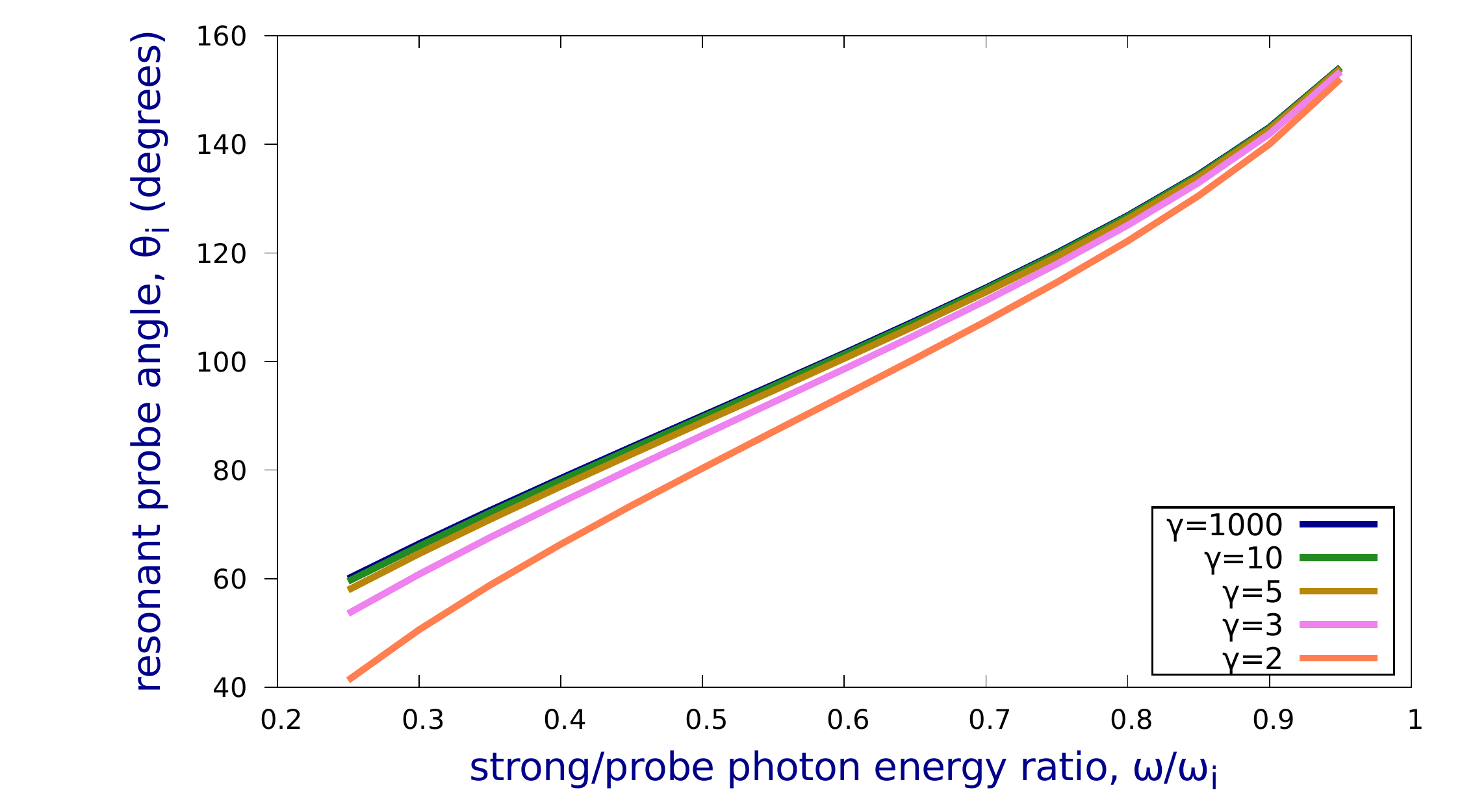}\caption{\bf Resonant probe angle vs photon energy ratio for relativstic $\gamma$ electrons.}
\label{fig:theti}\end{figure}

\begin{gather}
\cos\theta_i\approx\mfrac{1+\beta +(1+\beta)^2l\omega/\omega_i +\xi^2/2\gamma^2}{\beta(1+\beta)-\xi^2/2\gamma^2} \quad \text{direct channel}\notag\\
\cos\theta_f\approx\mfrac{1+\beta -(1+\beta)^2l\omega/\omega_f +\xi^2/2\gamma^2}{\beta(1+\beta)-\xi^2/2\gamma^2} \quad \text{exchange channel}
\label{eq:resangles}\end{gather}

\begin{figure}
\centering
\includegraphics[width=0.8\textwidth]{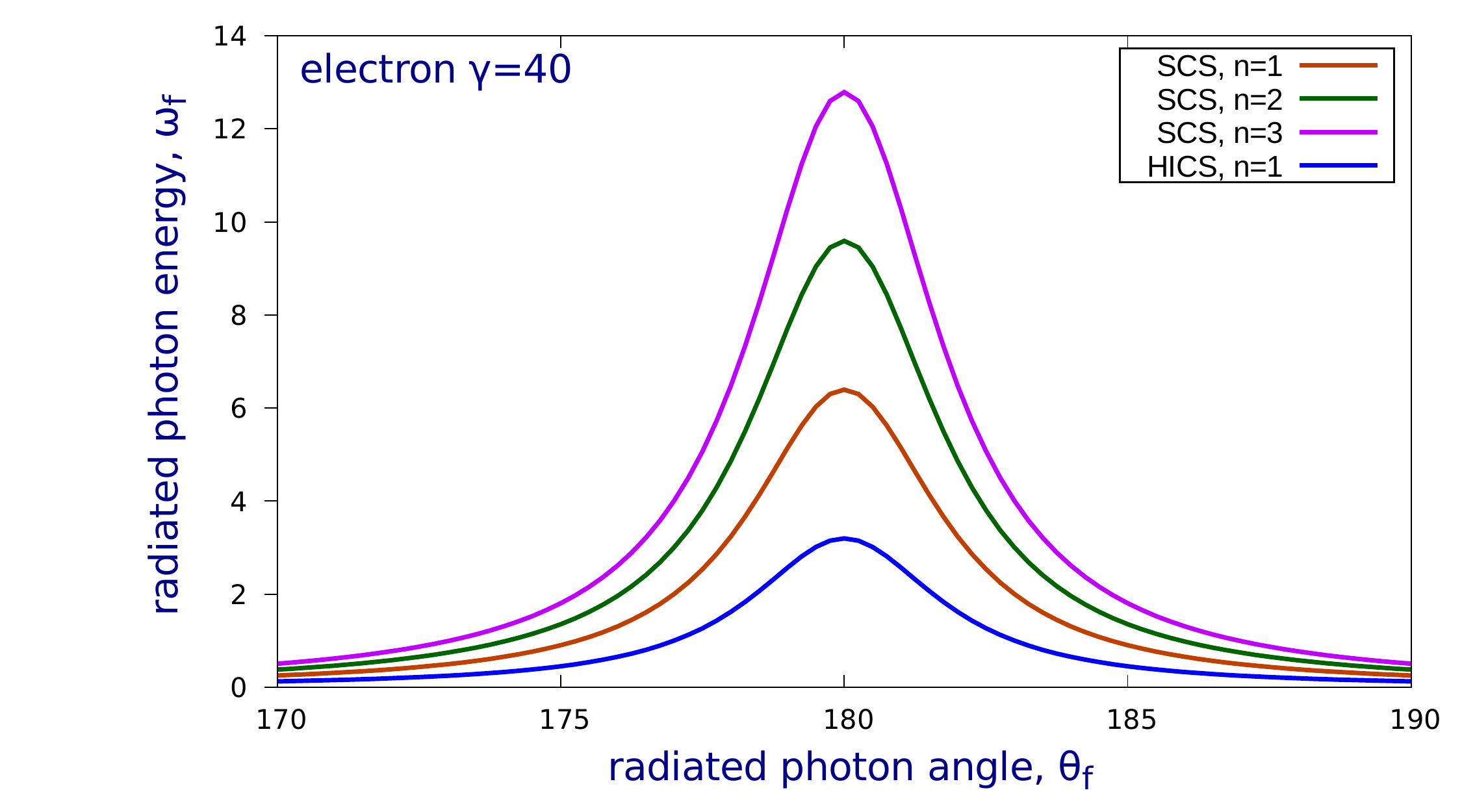}\caption{\bf Radiated photon energy vs radiation angle at resonance.}
\label{fig:thetf}\end{figure}

The direct channel resonance condition depends only on initial state parameters. Assuming $\xi\approx 1$ for the strong laser, and a broad range of electron energies, the tuning of the initial probe photon energy results in the resonant incident angles shown by figure \ref{fig:theti}. The energy of the radiated photons at resonance is given by the conservation of energy-momentum (equation \ref{eq:scswf}). The resonant radiated photons are smoothly distributed in a forward cone around the initial electron momentum. The energy of the radiated photon steps up for each external field mode corresponding to the overall contribution from the external field $nk$ (figure \ref{fig:thetf}). \\

\begin{gather}
\omega_f=\mfrac{(n+1)\,\omega\,\gamma(1+\beta)}{\gamma(1\!+\!\beta\cos\theta_\text{f})\!+\!\ls n\mfrac{\omega}{m}\!+\!\frac{\xi^2}{2\gamma(1+\beta)}\rs(1\mhy\cos\theta_\text{f})+\mfrac{\omega_\text{i}}{m}\!\ls1\mhy\cos(\theta_\text{i}\!+\!\theta_\text{f})\rs}
\label{eq:scswf}\end{gather}

The exchange channel resonance, by contrast, will show resonant radiated photon angles. The overall resonance structure is a combination of direct and exchange channels in the complete SCS transition probability. The width and height of the resonances, whether they are scanned over angles or probe energy, is dependent on both the numerator of the transition probability and the imaginary part of the self energy correction appearing in the denominator \cite{Hartin06}. \\

An estimate of the transition probability at resonance can be made by comparing the tree level propagator denominator to the resonance width. The resonance width is determined from the imaginary part of the Furry picture electron self energy. For a strong laser intensity $\xi\approx 1$, the resonance width can be given approximately by a function of the QED coupling constant $\alpha$, the field intensity $\xi$ and the recoil parameter $\chi$\cite{BecMit76},

\begin{gather}
\Gamma\approx 0.29\,\alpha\,\chi\,\xi^{0.86}
\end{gather}

\begin{figure}[htb]
\centering
\includegraphics[width=0.8\textwidth]{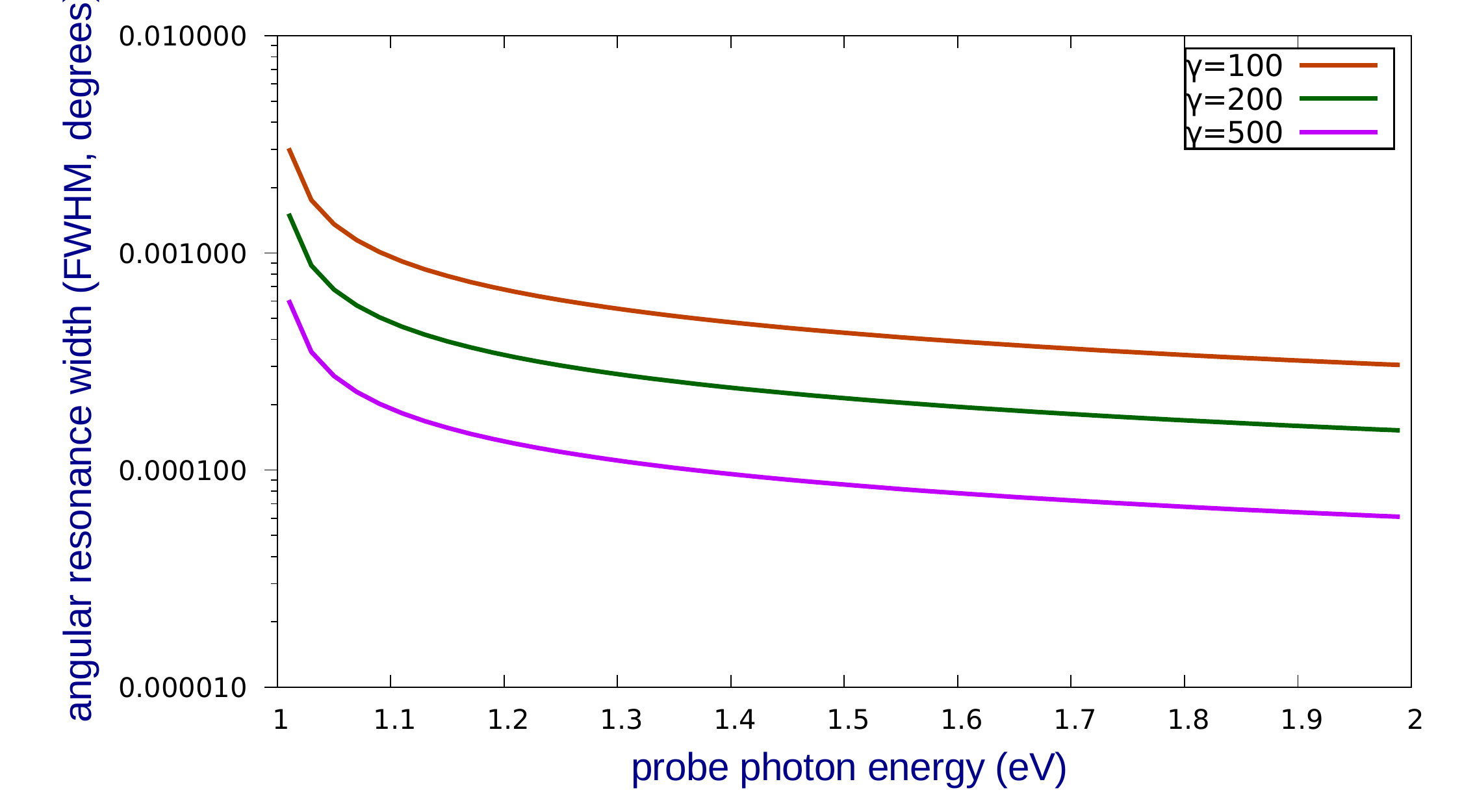}\caption{\bf Direct channel angular resonance width for 1eV laser photons}
\label{fig:fwhm}\end{figure}

The resonance width yields the angular resolution of the differential cross-section resonance using the expressions of equation \ref{eq:resangles}. Using the condition for the direct channel $l=1$ resonance, the peak angular resolution (FWHM) is up to 0.003 degrees for a 200 MeV electron, when the ratio between probe photon energy and laser photon energy approaches unity (figure \ref{fig:fwhm}). The resonance broadens as the electron energy decreases. For higher electron energies, the angular resolution decreases towards the limit of what was experimentally resolvable in similar past experiments \cite{McDonald91}. Correspondingly, a small angular resolution will give a precise measurement of the resonance location. \\

In a real experiment, there will be other factors that will play a role. Specifically, the emittance of the electron beam and the fact that the strong laser field will be provided by a pulse, are factors that will broaden SCS resonances. Other factors will be the resolution of detectors and background processes. Nevertheless, the magnitudes are such\cite{Hartin06} that a dedicated experiment should be able to distinguish these new predicted resonant effects. \\

We turn now to a different arena where strong field physics manifests itself.

\section{Strong field effects in colliders}\label{sect:colliders}

\subsection{Theoretical approaches}\label{sect:collapproach}

Since particle colliders collide dense charged particle bunches, intense electromagnetic fields are present at the point of collision. Strong field effects in colliders have been employed to explain the beamstrahlung \cite{YokChe91}, depolarisation \cite{SokTer64}, coherent pair production \cite{JacWu89,ChePal92} and higher order effects \cite{Chen92} at the interaction point (IP). \\

The Furry picture has not been the only method used to explain the strong field beam beam effects. For instance incoherent pair backgrounds rely on the Weizs\"acker-Williams equivalent photon approximation, \cite{Weizs34,William35} combined with the normal (non Furry picture) QED perturbation theory. This approach results in a variety of pair production processes, which are categorised depending on whether the electromagnetic fields of one, both or none of the colliding bunches are rendered as equivalent photons \cite{Schulte99,Yokoya03}.\\

Another approach is to consider the relativistic motion of the colliding fermions as classical and their interactions as quantum. Transition probabilities are calculated by this quasi-classical operator method (QOM) with relativistic quantum mechanics. This approach has proved successful in predicting diverse phenomena such as beamstrahlung rates \cite{Baier69}, the anomalous magnetic moment in a background field \cite{BaierKS76} and equivalent channeling phenomena in crystals \cite{Ugger05}. However, the QOM requires that all charged particles involved in a process be ultra relativistic. Its validity is questionable for the beamstrahlung process when a large energy photon is radiated, thereby rendering the final state non ultra relativistic. \\

This review however, will concentrate on analyses using the Furry picture (FP) framework within non perturbative QFT. It has been shown, for the case of the beamstrahlung at least, that the transition probabilities obtained using the FP and the QOM are asymptotically identical in the ultra relativistic limit \cite{Hartin09}. However, the FP does not require an ultra relativistic limit and is exact for all kinematics. \\

Collider processes can in principle all be studied in the Furry picture. Though there is still much work to be done, studies have been performed for background processes \cite{Hartin13a}, spin tracking processes \cite{Hartin11d}, precision processes \cite{Hartin15} and higher order resonances \cite{Hartin06}. \\ 

The Volkov solution of the Dirac equation in a constant crossed field is ubiquitous in Furry picture treatments of collider IP processes. More comprehensive analyses, especially for precision strong field processes, can make use of solutions in the fields of both colliding bunches \cite{Hartin15}. This review proceeds by considering first the collider strong field parameters.

\subsection{Strong field parameters at colliders}

The strength of the electromagnetic field at the IP of a collider is an important parameter in determining the transition probability of FP processes. The more relativistic the colliding particles are, the stronger the electromagnetic fields appear. \\

To set a numerical scale for an electromagnetic field strength which leads to detectable effects, the Schwinger critical field of $1.3\times10^{18} \text{ V m}^{\mhy2}$ can again be employed. The field strength of the intense charge bunches colliding at the IP, in the rest frame of oncoming particles, is expressed as a ratio to the Schwinger critical field, denoted by the so-called $\Upsilon$ parameter. The onset of intense field effects at the IP is $\Upsilon\gtrsim0.1$. A whole range of non-linear, intense field effects come into play, only a few of which have been rigorously studied. \\

In a real bunch collision, the bunch distorts due to electromagnetic effects, leading to variation in $\Upsilon$ at each point where a particle process takes place. An average value $\Upsilon_{\text{av}}$ can be defined across a whole bunch collision. The $\Upsilon_{\text{av}}$ parameter for a particular collider depends on the interaction point beam parameters - the bunch population N, the bunch dimensions $\sigma_x,\sigma_y,\sigma_z$ and the bunch particle energy (expressed as the relativistic $\gamma=E/m$), as well as the Compton radius $r_e$ and the fine structure constant $\alpha$\cite{YokChe91},

\begin{gather}\label{equpsav}
\Upsilon_{\text{av}}=\mfrac{5}{6}\mfrac{N\;\gamma\; r_e^2}{\alpha (\sigma_x+\sigma_y)\sigma_z}
\end{gather}

Using $\Upsilon_\text{av}$ as a guide, strong field parameter sets can be assembled for various colliders, both past and planned. In previous lepton colliders, $\Upsilon_{\text{av}}$ was vanishingly small so that intense field effects were not to be expected. However, in the next generation of $e^{+}e^{-}$ colliders (the international linear collider and the compact linear collider), $\Upsilon_\text{av}$ is in the strong field QED regime ($\Upsilon_{\text{av}}\gtrsim 0.1$, see table \ref{upsav}). \\

In hadron colliders like the LHC, $\Upsilon_\text{av}$ is very small due to the large rest mass of hadrons and therefore sluggish response to the IP electromagnetic fields, compared to the response of relatively light $e^{+}e^{-}$. Future $e^{+}e^{-}$ colliders will really provide the first occasion where bunch field strengths will be large enough to lead to significant non-linear effects. Only a precise theoretical calculation and simulation of these effects can tell us what to expect. \\

\begin{table}[h!]
\tbl{\bf Collider beam parameters and the strong field $\Upsilon_{\text{av}}$}{
\centering\begin{tabular}{|c|| c | c | c | c |}\hline
 Machine & LEP2 & SLC & ILC & CLIC \\ \hline\hline
   E (GeV) & 94.5 & 46.6 & 500 & 1500 \\ \hline
    $N(\times 10^{10})$ & $334 $ & $4$ & $ 2 $ & $ 0.37 $\\ \hline
   $\sigma_x,\sigma_y$ ($\mu$m) & 190, 3 & 2.1, 0.9 & 0.49, 0.002 & 0.045, 0.001\\ \hline
   $\sigma_z $ (mm) & 20 & 1.1  & 0.15 & 0.044\\ \hline\hline
   ${\bf \Upsilon_{\text{av}}}$ & 0.00015 & 0.001 & {\bf 0.24} & {\bf 4.9} \\ \hline
\end{tabular}\label{upsav}}\end{table}

Simulation of the IP bunch collision requires specific understanding of effects that lead to an increase in the electromagnetic field strengths. The main mechanism which increases the field strengths is the pinch effect, i.e. the effective focussing of each bunch in the field of the other \cite{YokChe91,Schulte99}.

\subsection{Beamstrahlung as a strong field Furry picture process}

The beamstrahlung is the photon radiation by charged particles in a collider charged particle bunch at the IP, due to the strong field of the oncoming charged bunch. The transition rate of the beamstrahlung process can be considered within the framework of the Furry picture in which the strong field of the oncoming bunch can be taken into account exactly. \\

The calculation proceeds in the same way as the calculation for the one vertex photon radiation in an intense laser, except the form of the external field is a constant crossed one instead of circularly, linearly or elliptically polarised. Indeed, it is relatively straightforward to develop the formalism for a general plane wave field and then substitute the precise form in at the end of the calculation \cite{Hartin11a}. \\

The transition probability $W_\text{beam}$ can be expressed in either McDonald's functions\cite{Baier09}, or equivalently, in Airy functions \cite{Ritus79},

\begin{align}\label{beamstr}
 W_\text{beam} &=\mfrac{\alpha m^2}{\pi\epsilon\sqrt{3}} \int_{0}^\infty \nths\mfrac{du}{(1+u)^2}\left[\int_{\chi}^\infty \nths K_{5/3}(y)dy-\mfrac{u^2}{1+u}K_{2/3}(\chi)\right] \quad \text{where} \quad \chi=\mfrac{2u}{3\Upsilon}
\end{align}

\begin{figure}[htb]
\centering
\includegraphics[width=0.8\textwidth]{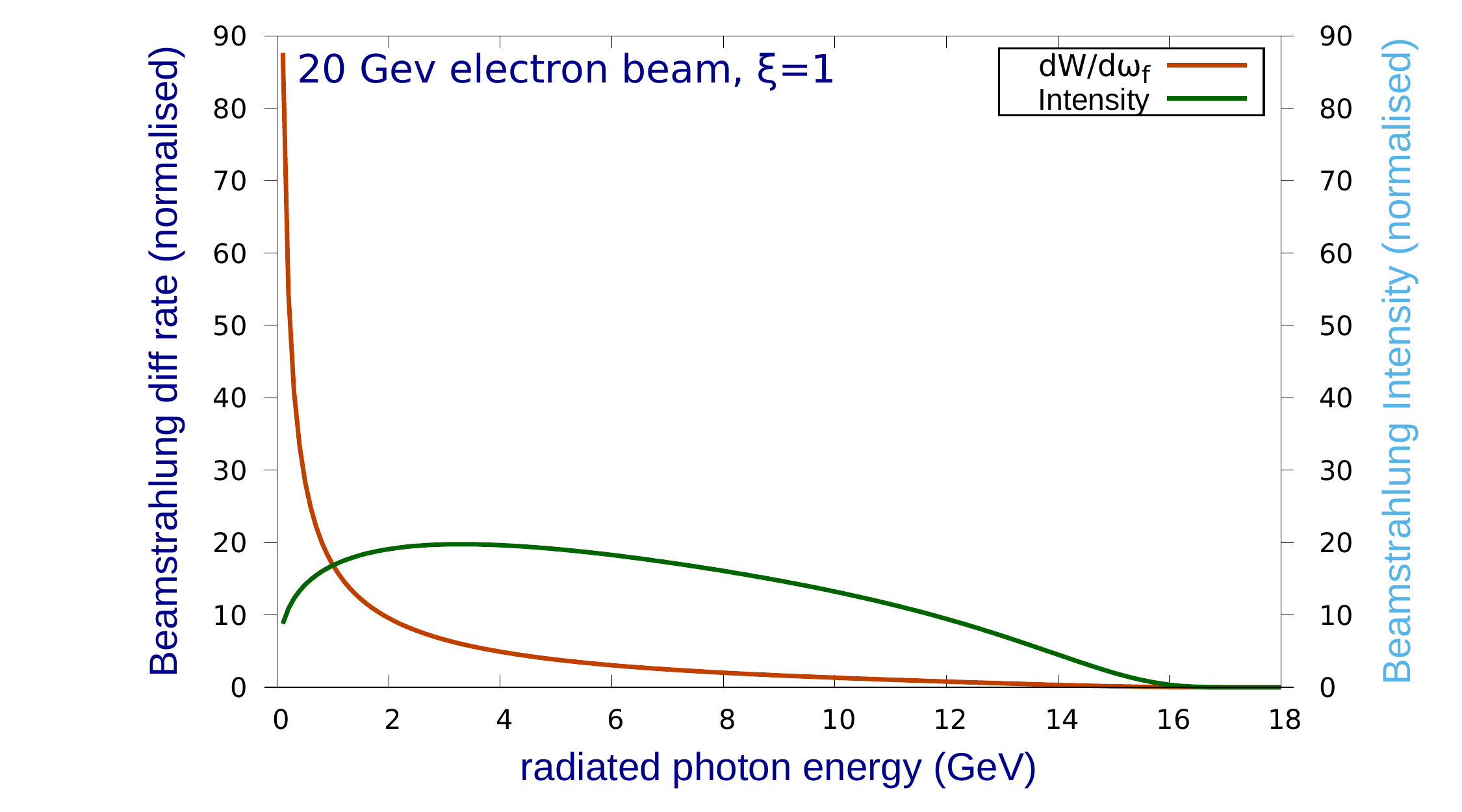}\caption{\bf Differential beamstrahlung rate and radiated photon intensity}
\label{fig:beamstr}\end{figure}

A correspondence of this quantum formula with the classical treatment is found by taking the limit of small external field intensity. In this limit, the second term of equation \ref{beamstr} vanishes. The remaining term diverges at the lower limit of the integration. This infrared divergence can be handled by including additional self-energy terms\cite{Hartin11b}. In practice, the transition probability is often given as an radiation intensity, which remains finite at the low radiation energy limit (figure \ref{fig:beamstr}), or a low energy cut at the limit of a real detector resolution is introduced.

\subsection{Strong field collider pair production processes}

The beamstrahlung is a copious source of high energy photons which can themselves interact with the oncoming bunch field in further strong field processes. Usually, these pair background processes are divided into coherent and incoherent processes, depending on how the external field is treated. \\

In the Bethe-Heitler process \cite{BetHei34} a real beamstrahlung photon combines with an equivalent photon from the Weiszacker-Williams treatment of the oncoming field to produce the background pair. In the Landau-Lifshitz process, \cite{LanLif34} both initial photons are derived from the equivalent photon approximation. The Breit-Wheeler process \cite{BreWhe34} combines two actual beamstrahlung photons.

\begin{align}
\text{\bf Incoherent pair processes}&\notag\\
\text{Breit Wheeler process }\quad \gamma_\text{real}+\gamma_\text{real} &\rightarrow e^{+} + e^{-} \notag\\
\text{Bethe Heitler process }\quad \gamma_\text{real}+\gamma_\text{equiv} &\rightarrow e^{+} + e^{-} \notag\\
\text{Landau Lifshitz process }\quad \gamma_\text{equiv}+\gamma_\text{equiv} &\rightarrow e^{+} + e^{-} \\[8pt]
\text{\bf Coherent pair processes}& \notag\\
\text{One photon pair production }\quad \gamma_\text{real}+n\gamma_\text{field} &\rightarrow e^{+} + e^{-} \notag\\
\text{Stimulated two photon pair production }\quad \gamma_\text{real}+\gamma_\text{real}+n\gamma_\text{field} &\rightarrow e^{+} + e^{-} \notag\\
\text{Trident pair production }\quad e^{-}+n\gamma_\text{field} &\rightarrow e^{-}+e^{+} + e^{-} \notag
\end{align}

The coherent pair production processes in colliders are equivalent to strong field processes in strong laser fields, except with the external field given by the constant crossed field of the oncoming charge bunch\cite{Yokoya03}. Generally, only the one photon pair production (OPPP) process has been fully taken into account\cite{Schulte99,Yokoya03}. However, the Breit-Wheeler process itself occurs in the strong field of the oncoming bunch \cite{Hartin07}. It should be treated as a resonant strong field process, the stimulated two photon pair production (figure \ref{fig:2ndorder}). \\

The resonant Breit-Wheeler process at the collider IP was calculated theoretically by inclusion of the electron self-energy in a constant crossed field as a resonance width\cite{Hartin06,Hartin11b}. A numerical calculation for parameters expected at future linear colliders shows that the differential transition rate at resonance, can be far in excess of those of the one vertex OPPP coherent process \cite{Hartin06}. \\

As replicates the situation with strong laser fields, a full calculation of the trident process at colliders is still under way. Often, it is simply treated as the two step process with an on-shell intermediate photon. The one step trident process is also potentially resonant, after due consideration of the finite pulse length of the charge bunch field (section \ref{sect:1steptrid}). \\

It is important to appreciate that higher order resonant effects were not likely to be seen at previous lepton colliders due to the relatively low field value (i.e. low $\Upsilon$ parameter). The onset of the strong field regime at future linear collider interaction points demands that such higher order resonant effects be simulated in a full PIC code of the beam-beam interaction. Such efforts are under way in a dedicated software package, {\bf IPstrong} (section \ref{sect:sim}).
 
\subsection{Spin dependent strong field interactions at the collider interaction point}

The precision physics program of future linear colliders requires polarised lepton beams and a careful accounting of depolarisation processes \cite{Hartin11d}. Due to the strength of the IP fields, there is significant depolarisation at that point. A careful consideration, requires analytic calculation of spin dependent strong field processes \cite{Hartin11b}. \\

The spin dependent strong field processes arising from unwanted background processes include the Sokolov-Ternov spin flip (section \ref{sect:spinflip}). The classical precession of the particle spin is also a significant contributor to depolarisation. The spin precession includes the anomalous magnetic moment, which is itself a strong field Furry picture process since it is in the presence of the colliding bunch fields  (section \ref{sect:precess}). \\

Higher order, resonant, strong field processes are also spin dependent. Since the resonant rates for these processes can be large, any thorough consideration of strong field spin dependence would study these processes as well. Such a calculation requires higher order Furry picture helicity amplitudes. New methods are available to make the analytic work tractable\cite{Hartin16}. \\

Any simulation of spin dependent strong field processes will have to take into account the spatial offset of the colliding beams. Such offsets can lead to regions of very intense fields and increase the overall depolarisation. The offsets of colliding beams, among other things is governed by misalignment due to environmental noise (section \ref{sect:sim}). \\

Finally, all processes that occur at the interaction point are strong field processes. For precision processes that are used to test theoretical models, the spin dependence on the strong field background is highly relevant. One such process is the strong field W boson pair production (section \ref{sect:wpair}).

\subsection{Spin precession and the strong field anomalous magnetic moment} \label{sect:precess}

The spin precession is classical but strong field quantum effects enter through the anomalous magnetic moment (AMM) in the charge bunch field. Spin precession is described by the Thomas-Bargmann-Michel-Telegdi (T-BMT) equation and describes the time evolution of the fermion spin vector $\vec{S}$ under the influence of a transverse and longitudinal magnetic field $\vec{B}_\text{T},\vec{B}_\text{L}$ and an electric field $\vec{E}$

\begin{align}\label{tbmt}
\mfrac{{\rm d}\vec{S}}{dt}=-\mfrac{e}{m\gamma}\ls(\gamma a_e+1)
\vec{B}_\text{T}+(a_e+1)\vec{B}_\text{L}-\gamma\lp a_e+\mfrac{1}{\gamma+1}\rp\beta
\vec{e}_v\times \mfrac{\vec{E}}{c}\rs\times \vec{S}
\end{align}

A strong field treatment for the T-BMT equation requires that the AMM ($a_e=\frac{g-2}{2}$) be replaced with an upsilon dependent expression, $a^\text{FP}_e(\Upsilon)$ which can be obtained by calculating the spin dependent part of the mass operator using exact Volkov solutions \cite{Ritus70} or by the QOM\cite{BaierKS76}. The result for a constant crossed field is an integration over a Scorer function\cite{Nist10},

\begin{gather}
a^\text{FP}_e=\mfrac{\alpha}{\sqrt{\pi}}\int_0^{\infty}\mfrac{du}{(1+u)^3} \,z\, \text{Gi}(z) ,\quad\text{where}\; z=\lp\mfrac{u}{\Upsilon}\rp^{2/3}
\end{gather}

\begin{figure}[h!] 
\centerline{\includegraphics[width=0.65\textwidth]{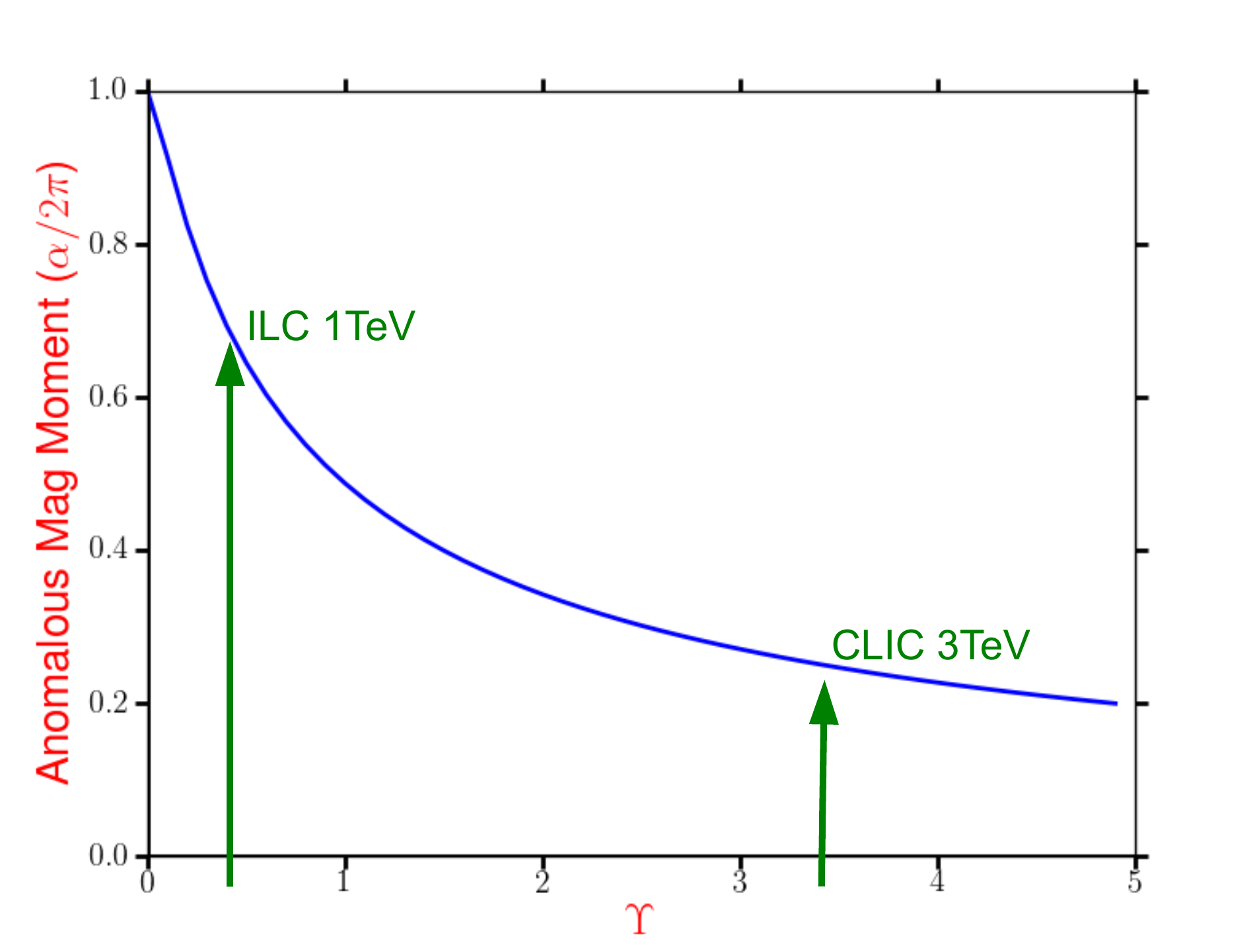}}
\caption{\bf The variation of the electron anomalous magnetic moment in a background field.}\label{fig:FPamm}
\end{figure}

When plotted (figure \ref{fig:FPamm}), the AMM in the external field declines from it's one loop value of $\frac{\alpha}{2\pi}$\cite{Schwinger48a} as the strength of the field increases. \\

Higher order corrections to $a_e$ can be obtained using equivalent processes within the Furry picture. By and large, these higher order strong field contributions have yet to be calculated. \\

For experiments that measure the AMM of the muon $a_\mu$\cite{Bennett04}, it may prove important to use a Furry picture analysis for the theoretical calculations of it's value. The external fields may be small in any given experiment, but a small variation in the Furry picture value $a^\text{FP}_\mu$, has potentially a large impact, given that it constitutes one of the largest anomalies indicating physics beyond the standard model.

\subsection{The spin dependent beamstrahlung and interaction point depolarisation} \label{sect:spinflip}

The spin dependent beamstrahlung process was derived historically through a quantum treatment of the problem of synchrotron radiation \cite{SokKleTer52}. With inclusion of polarisation and spin parameters, it was realised that the rates were spin dependent \cite{Ternov95,SokTer64}. Different calculation methods included the quasi-classical operator method \cite{Baier68,BerLifPit82} and the Furry picture \cite{Ritus72}. \\

The Sokolov-Ternov (S-T) effect is a well known feature of storage rings in which transverse polarisation of the circulating beam builds up over macroscopic time scales. Here however, we are interested in the same process in the more shorter time span of the charge bunch collision but with much larger field strengths. \\

Following the Furry picture treatment, the transition probability $W_\text{beam}$ of the radiation of a polarized electron of momentum $p$ with spin vector $s_\nu$ and Airy function arguments, $z$ in a strong constant crossed field is \cite{Ritus72}

\medskip
\begin{align}\label{polbeamstr}
W_\text{beam} &=\mfrac{\alpha m^2}{\pi\varepsilon_\text{p}} \medint\int_{\nthn ~0}^\infty \!\mfrac{du}{(1+u)^2} \ls \Ai_\text{1}+\mfrac{2+2u+u^2}{z(1+u)}\Ai'-\mfrac{eF^{\ast\mu\nu}p_\mu s_\nu}{m^3(1+u)} z\text{Ai}\rs,\,z=\ls\mfrac{u}{\chi}\rs^{2/3}
\end{align}

The final, spin dependent term in the beamstrahlung transition rate is what leads to the depolarisation. Taking into account possible operating parameters of future linear collider designs (table \ref{tab-pac11-1}), the calculated depolarisation contributions from the various spin dependent processes can be calculated (table~\ref{tab-pac11-2}), using appropriate simulation programs\cite{Hartin11d}.

\begin{table}[h]
{\footnotesize
\tbl{\bf Parameters sets for possible future linear collider designs.\label{tab-pac11-1}}
{\begin{tabular}{|l|c|c|c|}
\hline
 & Set 1 & Set 2 & Set 3\\ \hline
$\sqrt{s}$/GeV & 500 & 3000 & 3000\\
N $/10^{10}$  & 2  & 0.37 & 0.37\\
$n_B$        & 2625 & 312 & 312\\
$\gamma \epsilon^{*}_x$/mm mrad & 10 & 0.66 &  0.66\\
$\gamma \epsilon^{*}_y$/mm mrad & 0.04 & 0.02&  0.02\\ 
$\beta^{*}_x$/mm & 20 & 4.0 & 6.9\\
$\beta^{*}_y$/mm & 0.4 & 0.09& 0.068\\
$\sigma_z$/$\mu$m & 300 & 45 & 44\\
${\cal L}_{99 \%}$/$10^{34}$cm$^{-2}$s$^{-1}$ & 2.0 & 2.0 & 2.0\\ \hline
\end{tabular}}}
\end{table}

\begin{table}[h]
\tbl{\bf \hspace{1cm} Comparison of the luminosity-weighted depolarising effects in beam beam interactions for possible future linear collider parameter sets with fully polarised incident beams. T-BMT (S-T) denotes  effects due to spin precession (synchrotron radiation).}
{\begin{tabular}{|l|c|c|c|}
\hline
Parameter set & \multicolumn{3}{c|}{Depolarization $\Delta P_{lw}$}\\ 
      & Set 1         & Set 2  & Set 3\\ \hline
T-BMT &      0.17\% & 0.10\%   & 0.09\%    \\
S-T &        0.05\% & 3.40\%   & 3.81\%   \\
incoherent & 0.00\% & 0.06\%   & 0.00\%   \\
coherent &   0.00\% & 1.30\%   & 1.51\%   \\ 
total &      0.22\% & 4.80\%   & 5.53\%   \\ \hline
\end{tabular}\label{tab-pac11-2}}
\end{table}

With larger charge bunch field strengths, both the depolarisation and the contribution from coherent strong field processes, is more significant. This underlines the importance of performing higher order, strong field, spin dependent calculations in order to take into account all sources of linear collider IP depolarisation.

\subsection{Collider precision physics and strong field effects} \label{sect:wpair}

Since coherent background processes at future linear colliders are not insignificant, the effect of strong bunch fields on precision physics processes should be looked at more closely. That means considering higher order physics processes in the Furry picture (FP). As an example, the strong field, W boson pair production process is considered in this section.\\

A future lepton collider is intended to be a precision machine. A substantial part of the physics programme requires polarised beams with the polarisation known to 0.25\%. Much work has gone into the design of the polarimeters in order to achieve high precision on polarimetry measurements \cite{BooHar09}. \\ 

The real, luminosity weighted, polarisation must be an interpolation between the upstream and downstream polarimeter measurements, since the measurement upstream necessarily excludes the IP depolarisation and the measurement downstream is an over-estimation \cite{Moort08}. Cross-checks of the polarimetry interpolation is obtained from cross-section data. \\

In particular, the $W^{+}W^{-}$ pair production can be used to calculate the luminosity weighted depolarisation up to a relative error of 0.2\% using the Blondel scheme \cite{Rosca13}. However, there are remaining theoretical uncertainties in the $W^{+}W^{-}$ pair production data, due to the intense bunch fields. These strong field effects must be quantified in order to be sure of the luminosity weighted polarisation \cite{Aurand08}. \\

Using the Furry picture, the decays of the W boson in the presence of a single constant crossed electromagnetic have been calculated\cite{Obukhov87,Kurilin04}. Such calculations make use of the Proca equation in the presence of a background field \cite{Obukhov84}. In the limit of vanishing bunch field this process reduces to that described by the usual perturbation theory \cite{BeeDen94}. \\

Additionally, there are likely to be spin-dependent intense field effects on the $W^{+}W^{-}$ pair production process which are the equivalent of the spin flip in the beamstrahlung. The extent of the spin dependent effects, and thus the uncertainty in the polarisation measurement of the beams at the IP, can only determined by an analytic calculation and simulation within a real bunch collision. \\

\begin{figure}[h!] 
\centerline{\includegraphics[width=0.35\textwidth]{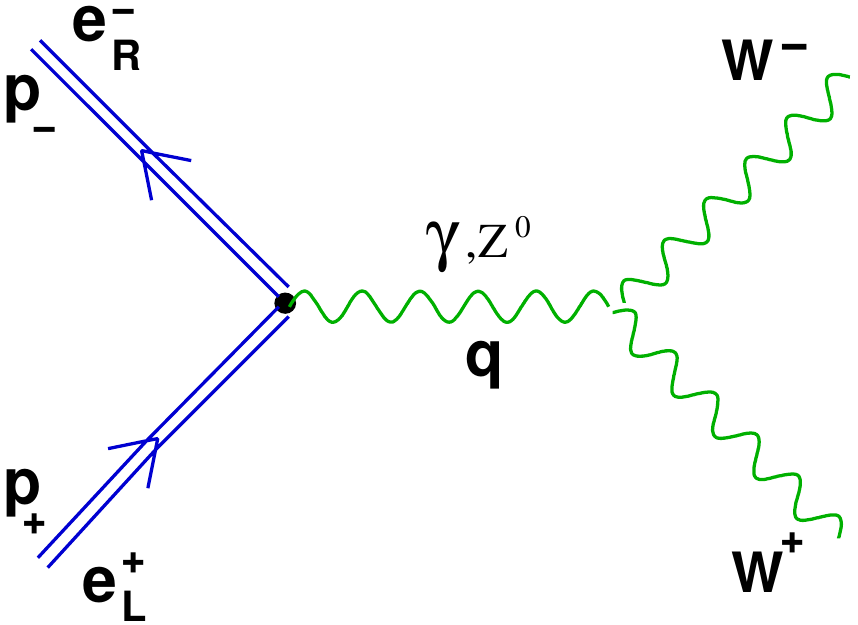}}
\caption{\bf The W boson pair production s channel.}\label{fig:feynwpair}
\end{figure}

The calculation is sketched here, outlining it's main features using the s-channel $W^{+}W^{-}$ pair production (figure \ref{fig:feynwpair}). Firstly, both initial and final particles couple to the background external field, so strictly speaking, both vertices are dressed by the Furry picture wave function for the leptons and gauge bosons.

In practice, the Furry picture wave function solutions are dependent on the strong field intensity parameter $\xi$ (equation \ref{eq:sqedparams}), which is proportional to the background field strength and inversely proportional to the mass of the particle that couples to that background field. As the field intensity parameter becomes small, the wave function becomes that of the free particle. \\

For the massive W boson, the field intensity parameter is much smaller than that for the initial $e^{+}e^{-}$ pair. As a reasonable approximation then, the Volkov solution for the leptons can be used together with the non external field W boson wave functions. That is, the first vertex is dressed by the external field and the second remains undressed. \\

As is usual for transition probabilities within the Furry picture, the square of a matrix element containing Volkov solutions will lead to summations over contributions from the external field at the dressed vertex. \\

\begin{figure}[h!] 
\centerline{\includegraphics[width=0.35\textwidth]{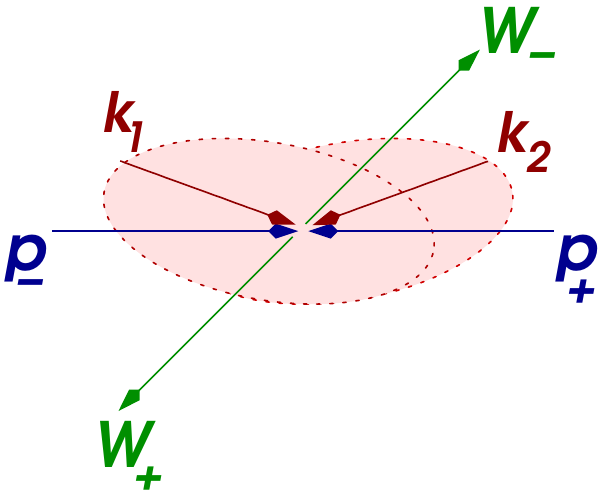}}
\caption{\bf The external field W boson pair production momenta. The photon momenta $k_1,k_2$ are associated with each of the colliding charge bunch fields.}\label{fig:wscatter}
\end{figure}

At the threshold of the $W^{+}W^{-}$ pair production, the gauge bosons will be produced with little longitudinal momentum. In that case the fields of both charge bunches appear to be of similar strength to the W bosons. In general, there should be allowance for a collision in which an incoming electron and positron, and the resulting produced pair, face an oncoming bunch field of differing momentum and field strength (figure \ref{fig:wscatter}) \\

%Each of the incoming particles sees the oncoming field of the other bunch. That is, the electron $p_{\!-}$ sees a field with 4-momentum $k_{\!+}$ and the positron $p_{\!+}$ sees $k_{\!-}$. Assuming that the beams meet head on in the CM frame, the kinematics of the initial states of the W pair production are, in terms of the beam energy $E$ and external field energy $\omega$

%\begin{gather*}
%p_{\!-}=(E,0,0,\sqrt{E^2-m_e^2}),\quad k_{\!+}=\omega(1,0,0,-1)\equiv\omega n_{\!+} \notag\\
%p_{\!+}=(E,0,0,-\sqrt{E^2-m_e^2}),\quad k_{\!-}=\omega(1,0,0,1)\equiv\omega n_{\!-}
%\end{gather*}

Instead of the usual Volkov solutions, exact solutions for both charge bunch fields should be used\cite{Hartin15}. These will lead to two summations $r,s$ over contributions from both charged fields at the dressed vertex. The two fields have intensity, momentum and potential $(\xi_\text{1},k_\text{1},A^e_\text{1})$ and $(\xi_\text{2},k_\text{2},A^e_\text{2})$ respectively. These all appear in the conservation of energy-momentum. \\

Writing the dressed vertex explicitly, including helicity states to make a spin dependent calculation, transforming into momentum space and specifying the final vertex $\gamma_\text{W}$, the transition probability has the following form, \\

\begin{gather}\label{Eq:FPtranprob}
W_\text{WPAIR}=\sum_{rs} |M_{fi}|^2 \;\mfrac{d\vec{W}_{\!-}d\vec{W}_{\!+}}{4\omega_{-}\omega_{+}}
\delta(W_{\!-}+W_{\!+}-p_{\!-\!}-p_{\!+\!}+r\xi_{1}\hat{k}_{1}+s\,\xi_{2}\hat{k}_{2}) \\
iM^e_{fi}= \int\! dr \;\mfrac{ie^2}{(p_{\!-}\!+\!p_{\!+}\!+\!rk)^2}\; 
\times \bar v_{p_{\!+}}\ls\!1-\mfrac{e\st{A}^e_1\st{k}_1}{2(k_1\!\cdot \!p^{\!+})}\!\rs\gamma^{\mu}\ls\!1-\mfrac{e\st{A}^e_2\st{k}_2}{2(k_2\!\cdot \!p^{\!-})}\!\rs\lp\mfrac{1\!+\!\gamma^5}{2}\!\rp u_{p^{\!-}}\,\gamma_\text{W} \notag
\end{gather}

Of course, since this is a second order Furry picture process, the resonances of the strong field propagator have to be taken into account. The propagator can be either a photon or a Z boson and both couple to the fields of both charge bunches through their self energies. \\

These are some of the main issues involved in carrying out a strong field calculation of general collider processes within the Furry picture. There is much more detailed analytic work to be done. Additionally, for the purposes of event generation and future detection of such effects, a detailed and dedicated simulation program must be constructed.

\section{Simulation of strong field physics at $e^{+}e^{-}$ and laser/electron colliders}\label{sect:sim}

The point of studying strong field physics processes theoretically, is to test the predicted phenomena experimentally. The glue that binds the experiment and theory together are simulations based on the theory targeted for particular experiments. \\

Some first order Furry picture (FP) processes (beamstrahlung and coherent pair production) are simulated in existing computer programs, which are dedicated to collider strong field effects. Notably, among these programs are CAIN \cite{Yokoya03} and GUINEA-PIG \cite{Schulte99}, which embed a monte carlo method within a Particle-in-cell (PIC) model of the beam beam interaction. There have been additional recent efforts to simulate the first order Furry picture processes in electron/laser interactions and laser/plasma interactions\cite{Ridgers14,Green15,Zhang15,Gonos15}. \\

These existing programs do not contain the structure to simulate higher order strong field processes such as the Furry picture $W^{+}W^{-}$ pair production in overlapping bunch fields (section \ref{sect:wpair}), or the resonant second order stimulated Compton scattering (section \ref{sect:scs}). However, the basic model of a monte carlo of the strong field transition probabilities, embedded in a PIC simulation, serve as a model for constructing a new, more comprehensive program, named {\bf IPstrong}\cite{Hartin18c}.\\

The primary purpose of {\bf IPstrong} is event generation of strong field processes in a general interaction involving one or more laser pulses, background fields, positively and negatively charged lepton bunches. The program structure is designed for future expansion to include hadron bunches, plasma and FEL interactions. {\bf IPstrong} will simulate equally well, first order as well as higher order Furry picture processes. \\

Project requirements primarily consist of event generation via a monte carlo method applied to Furry picture transition probabilities. For collider physics, the field strengths of both charge bunches serve as input into transition probabilities. To simulate a FEL interaction, it is crucial to include ponderomotive forces. For electron/laser interactions, Raleigh length and pulse shape should be taken into account. In all cases an accurate electromagnetic solver is applied on a real time basis throughout the simulation of a charge bunch collision. \\

Since computing is potentially intensive, a recent, object oriented version of Fortran is utilised\cite{fortran09}. The code is parallelised using Open MPI and currently computes the 3D electromagnetic field and charge evolution. The possibility of running the code on GPUs using CUDA\cite{cuda12} libraries is envisaged for future versions, with the initial version of the software running solely on CPUs. \\

In the interests of precision, the PIC electromagnetic simulation includes a 3D Poisson solver rather than previously used 2D solvers\cite{Yokoya03,Schulte99}. Charges are distributed to an adaptive grid that extends well beyond the actual charges in order to accommodate fringe fields. Output of events is in standard form, either in ASCII format for post processing or for STDHEP or LCIO format, for inclusion in geant4\cite{geant03} simulations of particle detectors\cite{ILCTDR13}. The particle pusher is a relativistic invariant version of a Boris pusher that includes ponderomotive forces. \\

\begin{figure}
%\centering\begin{subfigure}[t]{0.5\textwidth}
\centerline{\includegraphics[width=0.4\textwidth]{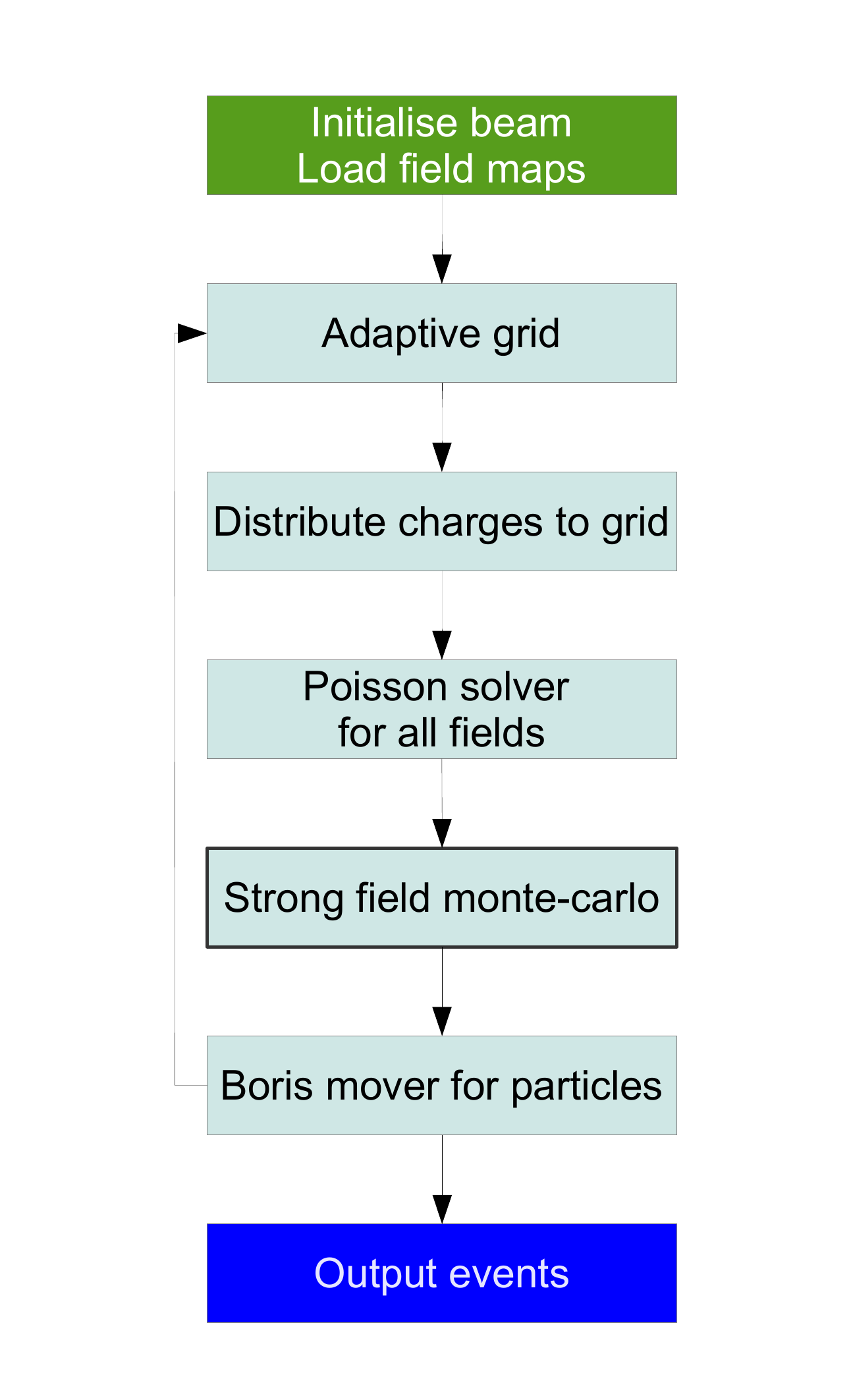}}
\caption{\bf Program modules and flow for simulation of IFQFT processes in a real bunch collision}\label{IPstrongflow}
%\end{subfigure}\begin{subfigure}[t]{.5\textwidth}
\end{figure}

{\bf IPstrong} implements a standard structure for strong field QED simulation programs, a monte carlo module embedded in a PIC code. The monte carlo relies on the acception/rejection method\cite{Yokoya03,robert16} and requires transition probabilities for particular processes, to be transformed so that the probability is flat over a particular parameter of interest. Figure \ref{IPstrongflow} depicts the {\bf IPstrong} program flow. \\

\begin{figure}
\centerline{\includegraphics[width=0.9\textwidth]{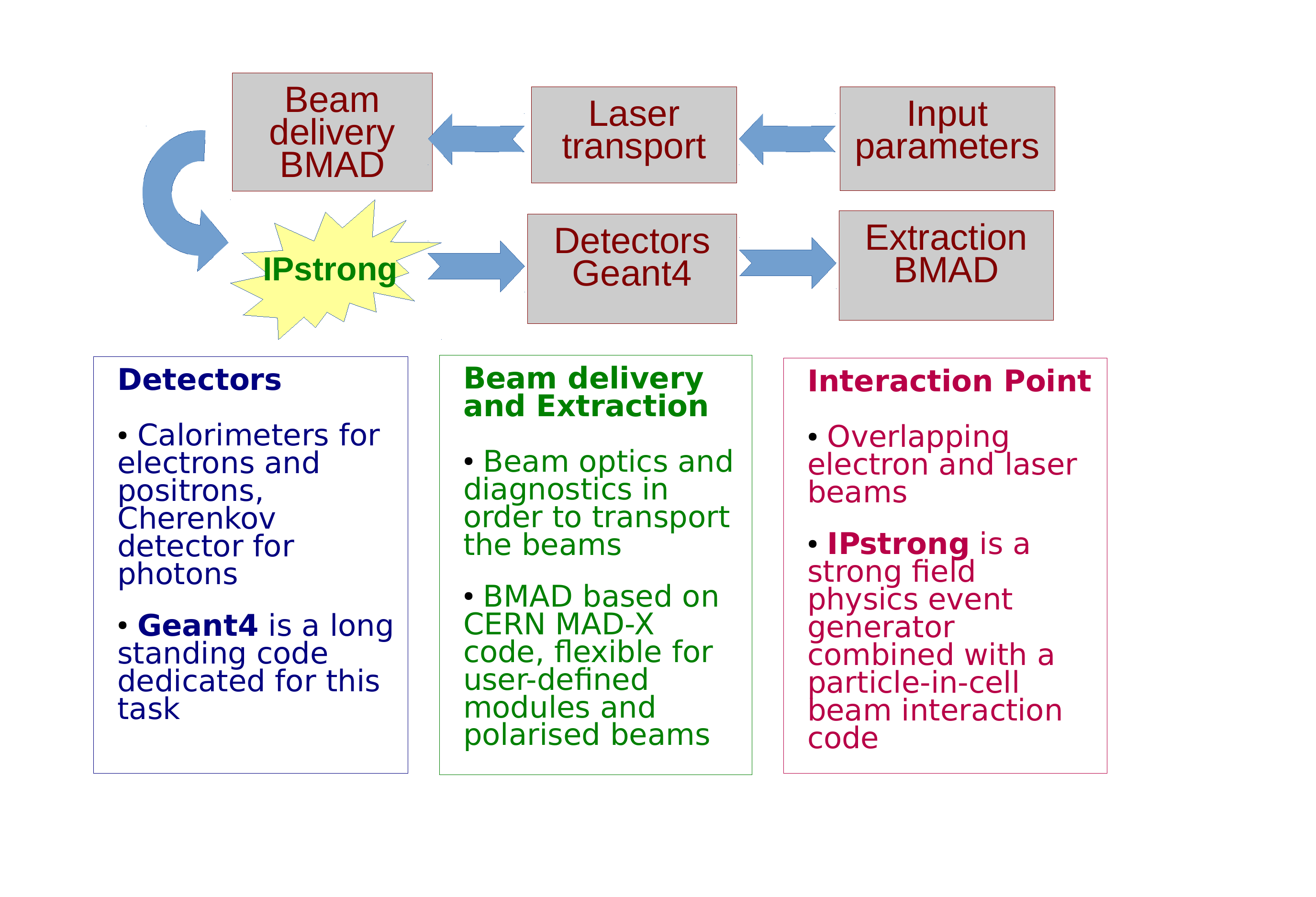}}
\caption{\bf Program modules and flow for simulation of beam delivery, interaction and detection for strong field IFQFT processes.}\label{fig:progflow}
%\end{subfigure}\caption{\bf Simulation modules and flow.}
\end{figure}

%Validation of the code, requires comparison of simulated trends with known analytic formula. Comparison with well established codes require the ability to work in electron/laser or lepton collider mode. Validation will take place alongside pre-existing codes such as CAIN and guinea-pig as well as contemporary codes that perform that same type of physics simulation. \\

For the purposes of cross-checking with existing programs, many features of a charge bunch collision such as the disruption angle, kink instability, pinch effect, centre of mass deflection and waist shifts are implemented \cite{YokChe91}. Full polarisation of the beams allows full investigation of the phase space for each macro-particle including energy, position, momentum and spin vector components. Strong field spin effects, such as the precession and Sokolov Ternov spin flip, as well as higher order spin effects, are also implemented for comparison with existing programs. Likewise, the electron/laser interaction will be validated against existing codes that perform the same type of physics simulations. \\

The actual bunch/pulse collision is simulated by taking time slices through the plane perpendicular to the group velocity of each bunch/pulse. Collision angles are implemented by Lorentz transformation to the anti co propagating frame, performing the monte-carlo and then transforming back to the original frame. As the colliding bunch/pulse slices overlap, a simulation grid is formed and the Poisson equation is solved on a cell by cell basis to determine the charge motion.\\

Since the starting point of a collision is important and is determined by pre-existing bunch or pulse evolution, IPstrong can be embedded in a wider program suite to take into account the realities of bunch/pulse delivery (figure \ref{fig:progflow}). For a real bunch collision occurring after passage through a beam delivery system, the final relative spatial offset of the bunches and their polarisation state are essential for intense field physics. \\

\begin{figure}[htb]
\centering
\includegraphics[width=0.7\textwidth]{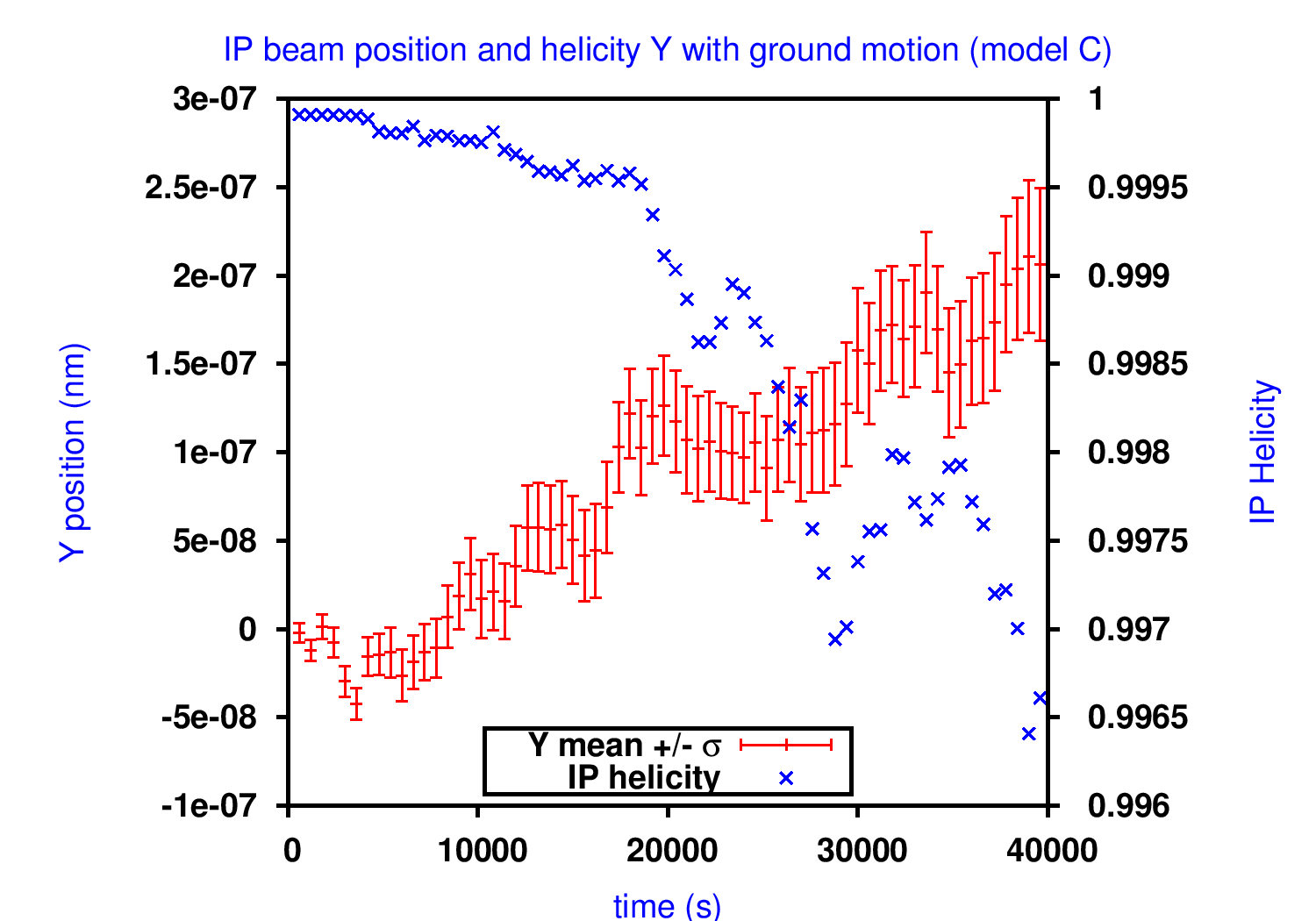}
\caption{\bf Ground motion induced beam offset and depolarisation for the ILC beamline with a moderate ground motion model.}
\label{fig:misalign}\end{figure}
%\end{comment}   %%%%%%%%%%%%%%%% ground motion %%%%%%%%%%%%%%%%

Any strong field experiment requires particle detectors and beam diagnostics\cite{Hartin10b} in order to measure final states. Typically, electron and positron detectors are calorimeters composed of, for example, alternating layers of tungsten and silicon, divided up into cells\cite{Bamber99}. Geant4 models can be included within the larger program suite. \\

An appropriate beam dynamics program, such as Bmad \cite{bmad06} simulates the transport of beams to and from the interaction point, as well as simulating beam diagnostics. One advantage of BMAD lies in its ability to transport spin, the components of which give new degrees of freedom in strong field phenomenology \cite{Hartin11d}. The effect of ground motion on beam jitter can be easily included. For instance, detailed studies of the misalignment of collider magnets due to ground motion showed that the bunches undergo displacement and depolarisation over time (figure \ref{fig:misalign})\cite{Hartin10b,Hartin11c}. \\

The coming period will see simulation programs like {\bf IPstrong} put to the test, as the strong field QED community look towards a new era of strong field QED experiments.

\section{Conclusions}

Quantum field theory with the non perturbative inclusion of strong electromagnetic fields, predicts interesting novel phenomena. The Schwinger critical field induces the creation of electron-positron pairs from the vacuum. For background fields of  intensity $0.1\!<\!\xi\!<\!1$, that can already be produced in the laboratory, higher order processes with a Furry picture propagator can reach the mass shell for physically measurable kinematics and should display large resonant transitions for tuned experimental parameters. These resonant transitions occur because of changes to the quantum vacuum brought about by a non perturbative inclusion of the background field. \\

These "vacuum resonances" can be understood in analogy to resonant atomic transitions. A quasi-energy level structure is set up in the vacuum by the strong electromagnetic field. The physical basis for these quasi-energy levels are likely to be the virtual charges themselves which respond to applied fields and preference certain virtual particle propagation states. \\

The existence of virtual charges is already known through well established phenomena like the Lamb shift. The Lamb shift, however is an atomic phenomena. The second order IFQFT processes outlined in section \ref{sect:res}, would manipulate vacuum charges using a circularly or linearly polarised electromagnetic field provided by a strong laser. \\

The location of resonances in parameter space is easily established by considering the momentum flowing into or out of the exchanged virtual particle. For tuned parameters, the resonances will appear as sharp peaks in scans over probe photon energy and angle of incidence. Given a careful selection of parameters, the experimental signals should be clear and their discovery would constitute a new test, both of QED and the non perturbative approach to it encapsulated by the Furry picture. \\

Indeed, the experimental tests outlined in this review, would be a test not just of QED, but of intense field quantum field theories, in general.  This class of theories predict a range of new phenomena including a shift in rest mass of particles which couple to the strong field. These theories predict also that the running of the coupling constant is modified. All of these predictions are amenable to experimental investigation and the current scope for new discoveries is large, given dedicated searches. \\

Even without probe photons, higher order resonant effects should be discernible in the simple interaction of relativistic electrons and an intense laser beam. The work horse process in this case is the one step trident process, which produces an electron, positron pair via a virtual, Furry picture photon. Since the virtual Furry picture photon couples to the strong field through a self energy loop, resonant transitions should be observed in scans of the final state momenta. \\

A full transition probability calculation uses Volkov solutions of the electron embedded in the external strong electromagnetic field. Recent work has simplified the procedure for obtaining the complete analytic expression, by adapting Fierz's method in order to swap the position of Volkov spinors in Furry picture matrix elements\cite{Hartin16}. \\

The standard approach in the Furry picture uses Volkov solutions for infinite plane wave electromagnetic fields. Additional modelling is required for conditions in the laboratory. Real, ultra-intense lasers are likely to be linearly polarised and pulsed. In this case, analyses which take into account the pulse length can be applied. In general, a finite pulse length smears out the conservation of energy momentum since the momentum state of external field momentum modes are not represented by delta functions, but some shape function. In the limit where laser pulses are long enough to contain many cycles, the Volkov solution is restored. \\

The additional Furry picture propagator poles, which correspond to transitions between quasi-levels, are regularised through the non perturbative self energy processes. These can be included in the propagator denominator via the formalism of lifetimes of unstable states. The physical picture which is suggested, is that the Furry picture electron and photon are unstable states which spontaneously undergo transitions in the dispersive vacuum. \\

If strong field effects are generated deliberately in electron/strong laser interactions, in lepton colliders the effects are involuntary. The collider interaction point involves dense charge bunches which interact whilst undergoing compression due to the pinch effect. The fields associated with these charge bunches can reach the Schwinger critical field in the rest frame of the interacting, relativistic particles. \\

In such circumstances, Furry picture processes abound at collider interaction points. The beamstrahlung itself is a Furry picture process, while large coherent pair backgrounds is another signature effect. In fact, {\bf all} primary collider processes occur in the intense fields of the interacting charge bunches. We should remind ourselves that the quantum vacuum is not electrically neutral and it polarises in the presence of the charge bunch fields. This must affect collider physics. \\

It will be particularly important to perform calculations of precision processes, such as the W boson pair production (section \ref{sect:wpair}), in the Furry picture. This is especially the case, since the higher order Furry picture processes have a dominant differential transition probability at strong field resonance. This is not just a nuisance, strong field effects in colliders may aid the search for new physics, not hinder it. \\

There are also implications for a theory of quantum gravity which formulates QFT in a curved background space\cite{Fewster08,BuchFrad81}. QFT in curved space times seems to be related to strong field QFT\cite{Marecki03} and any experimental confirmation of strong field QFT predictions, particular as regards its virtual particle or vacuum state, would be highly informative. To develop the correspondence between these theories, non perturbative solutions of the equations of motion on curved space-times or gravitational fields would be necessary\cite{BagObu92,Pollock10}, as well as propagators\cite{Goncal09} and particle processes\cite{Drummond80,DipCal06}. Technical difficulties remain, but theoretical programmes to overcome them exist\cite{Wald10}. \\

The current period in high energy particle physics is characterised by uncertainty about which direction nature goes beyond the standard model. The non perturbative, strong field theoretical predictions reviewed here, and their phenomenology in colliders and electron/laser interactions, promise to give new insights into the quantum vacuum and perhaps even indicate a way forward towards new physics. The challenge is to plan, simulate and carry out dedicated strong field experiments in the immediate future.

\section*{Acknowledgments}

The author would like to acknowledge funding from the Partnership of DESY and Hamburg University (PIER) for the seed project, PIF-2016-53. Additionally, this work was supported by a Leverhulme Trust Research Project Grant RPG-2017-143 and by STFC, United Kingdom.

\bibliographystyle{unsrt}
\bibliography{/home/hartin/Physics_Research/mypapers/hartin_bibliography}

\end{document}